\documentclass[AMS,Times1COL]{WileyNJDv5}

\makeatletter

\renewenvironment{appendix}{%
  \section*{APPENDIX}
  \global\appendixsectrue
  \setcounter{section}{0}
  \setcounter{figure}{0}
  \setcounter{table}{0}
  \setcounter{equation}{0}

  \renewcommand\thesection{\Alph{section}}
  \renewcommand\thefigure{\Alph{section}\arabic{figure}}
  \renewcommand\thetable{\Alph{section}\arabic{table}}
  \renewcommand\theequation{\Alph{section}\arabic{equation}}
}{}

\makeatother

\articletype{Research article} 

\usepackage{enumitem}

\usepackage{graphicx}
\usepackage{amssymb}
\usepackage{amsmath}
\usepackage{algorithm}
\usepackage{algpseudocode}
\usepackage{hyperref}

\usepackage{comment}
\usepackage{color}
\usepackage{mathtools}
\usepackage{multirow}
\usepackage{upgreek}
\usepackage{tikz}
\usepackage{enumitem}
\usepackage{makecell}

\usetikzlibrary{calc,patterns,
                 decorations.pathmorphing,
                 decorations.markings, 3d}

\usepackage{scalerel}

\newcommand{\R}{{\mathbb{R}}}
\newcommand{\N}{{\mathbb{N}}}
\newcommand{\E}{{\mathbb{E}}}

\newcommand{\unvec}{\mathrm{unvec}}
\newcommand{\I}[1]{\mathbb{I}_{[#1]}}
\newcommand{\tr}[1]{\mathrm{tr}\left(#1\right)}
\DeclareMathOperator*{\argmax}{arg\,max}

\newbool{arxiv}
\setbool{arxiv}{true}

\newcommand\setversion[1]{%
    \def\tempa{#1}%
    \def\tempb{student}%
    \ifx\tempa\tempb
        \setbool{arxiv}{true}%
    \else
        \def\tempb{teacher}%
        \ifx\tempa\tempb
            \setbool{arxiv}{false}%
        \else
            \errmessage{Unknown value for arxiv: #1}%
        \fi
    \fi
}

\newcommand{\mytitle}{From Data to Predictive Control: A Framework for Stochastic Linear Systems with Output Measurements}
\begin{document}

\title{\mytitle\thanks{This work has been supported by the Swiss National Science Foundation under NCCR Automation (grant agreement 51NF40 180545) and the ETH Career Seed Award funded through the ETH Zurich Foundation.}}



\author[IDSC,Harvard]{Haldun Balim}
\author[IDSC]{Andrea Carron}
\author[IDSC]{Melanie N. Zeilinger} 
\author[IDSC,ICL]{Johannes K\"ohler*} 
\authormark{Haldun Balim \textsc{et al}}
\titlemark{\mytitle}

\address[IDSC]{\orgdiv{Institute for Dynamic Systems and Control}, \orgname{ETH Zurich}, \orgaddress{\state{ Zurich CH-8092}, \country{Switzerland}}}
\address[Harvard]{\orgdiv{John A. Paulson School of Engineering and Applied Sciences}, \orgname{Harvard University}, \orgaddress{\state{Cambridge, MA 02138}, \country{USA}}}
\address[ICL]{\orgdiv{Department of Mechanical Engineering}, \orgname{Imperial College London}, \orgaddress{\state{London}, \country{United Kingdom}}}

\corres{Johannes K\"ohler \email{j.kohler@imperial.ac.uk}}


\abstract[Abstract]{We introduce data to predictive control, D2PC, a framework to facilitate the design of robust and predictive controllers from data.  The proposed framework is designed for discrete-time stochastic linear systems with output measurements and provides a principled design of a predictive controller based on data. The framework builds on a parameter identification method based on the Expectation-Maximization algorithm, which incorporates pre-defined structural constraints. 
An asymptotic approximation is leveraged to quantify the uncertainty in the parameter estimates. 
 As the main contributions, a robust control and predictive control design are proposed tailored to the uncertainty characterization resulting from the identification. 
In particular,  a strategy to synthesize robust dynamic output-feedback controllers is presented. Furthermore, a predictive control scheme that guarantees recursive feasibility and satisfaction of chance constraints is developed. This framework marks a significant advancement in integrating data-driven models into robust and predictive control designs. We demonstrate the efficacy of D2PC through a numerical example involving a $10$-dimensional spring-mass-damper system.
 }
\keywords{Model predictive control (MPC); robust control; stochastic MPC; data-driven control; identification for control}

\maketitle

\renewcommand\thefootnote{}
\footnotetext{\textbf{Abbreviations:} MPC, model predictive control; SMPC, stochastic model predictive control; LMI, linear matrix inequality; D2PC, data to predictive control; EM. Expectation-Maximization; MLE, maximum likelihood estimation; SOCP, second-order cone program}

 \begin{figure*}[htbp]
\centering
\begin{tikzpicture}

    \def \pcscale{4}
    
    \def \pcroboff{3.75}
    \def \pcuqoff{4}
    \def \tsdataoff{4}
    \def \dataidoff{2}
    \def \idscoff{4}
    
    \def \uqeh{.75}
    \def \firstcolheight{5}
    \def \firstcoloff{1.5}
    \def \dataaxsz{3}

    \def \scxoff{1.8}
    \def \scyoff{1.8}

    \def \ax{.2 * \pcscale}
    \def \ay{.2 * \pcscale}
    \def \eh{.15 * \pcscale}
    \def \ew{.05 * \pcscale}
    \def \aw{.8 * \pcscale}
    \def \ah{.6 * \pcscale}
    \def \textoffset{.15 * \pcscale}

    \def \begangle{10}
    \def \endangle{80}
    \def \ntubes{5}
    
    \draw[thick, ->] (\ax, \ay) +(0:\aw cm and \ah cm) arc (0:90:\aw cm and \ah cm);

    \foreach \i in {1,...,\ntubes} {
        \pgfmathsetmacro{\angle}{\begangle + (\endangle-\begangle) / (\ntubes-1) * (\i-1)}
        
        \pgfmathsetmacro{\ex}{\ax + \aw * cos(\angle)}
        \pgfmathsetmacro{\ey}{\ay + \ah * sin(\angle)}

        \draw[rotate around={90+\angle:(\ex,\ey)}, thick, fill=gray!20] (\ex, \ey) ellipse (\ew cm and \eh cm);
    }
    \node[align=center] at (\ax + \aw/2, \ay-\textoffset) {\textit{Section~\ref{sec:pred_control}:}\\Predictive Control};

    \huge
    \draw[thick, rounded corners] (\aw+\pcroboff, \ay+\ah/5) rectangle ++(4*\ah/5, 4*\ah/5) node[pos=.5, align=center] {$\Delta$};
    \normalsize
    \node[align=center] at (\aw+\pcroboff + 2*\ah/5 , \ay-\textoffset) {\textit{Section~\ref{sec:lqg}:}\\Robust Controller \\Synthesis};


    \def \elx{\aw+\pcroboff+4*\ah/5+\pcuqoff}
    \def \ely{\ay+3*\ah/5}
    \draw[fill=gray!20, thick] (\elx, \ely) ellipse (\uqeh*2 cm and \uqeh cm);

    \draw[dashed, -] (\elx - \uqeh *2 - .5, \ely) -- (\elx + \uqeh *2 + .5, \ely); 
    \draw[dashed, -] (\elx, \ely - \uqeh - .5) -- (\elx, \ely + \uqeh + .5); 
    \node[align=center] at (\elx , \ay-\textoffset) {\textit{Section~\ref{sec:unc_quant}:}\\Uncertainty\\Quantification};

    \draw[thick, rounded corners] (\elx-\ah/2, \firstcolheight) rectangle ++(\ah, \ah) node[pos=.5, align=center] {\textit{Section~\ref{sec:sysid}:}\\Parameter \\Identification};

    \def \tsw{4*\ah/5}
    \def \tsy{\firstcolheight + \ah/2 - \tsw/2}
    \draw[thick, rounded corners] (\ax+\firstcoloff, \tsy) rectangle ++(\tsw, \tsw) node[pos=.5, align=center] {True \\System};

    \def \axx{\ax+\firstcoloff+\tsdataoff}
    \def \yend{\firstcolheight+\dataaxsz*4/5}
    \draw[thick, ->] (\axx, \firstcolheight) -- (\axx+\dataaxsz, \firstcolheight); 
    \node[align=center] at (\axx+\dataaxsz+.2, \firstcolheight) {$t$};
    \draw[thick, ->] (\axx, \firstcolheight) -- (\axx, \yend); 
    \node[align=center] at (\axx, \yend+.2) {$y$};

    \def \xs{{0.1 , 0.13, 0.16, 0.19, 0.22, 0.25, 0.28, 0.31, 0.34, 0.37, 0.4 ,
        0.42, 0.44, 0.46, 0.48, 0.5 , 0.52, 0.54, 0.56, 0.58, 0.6 , 0.63,
        0.66, 0.69, 0.72, 0.75, 0.78, 0.81, 0.84, 0.87}}
       
    \def \ys{{0.60697754, 0.64691567, 0.71671434, 0.73777031, 0.71559555,
        0.76990041, 0.76133861, 0.6748491 , 0.65499159, 0.67364518,
        0.59749765, 0.57022917, 0.54419358, 0.51551787, 0.51277665,
        0.48637262, 0.42405463, 0.36401627, 0.3210516 , 0.27380235,
        0.30614852, 0.25113473, 0.19551048, 0.17965248, 0.15740828,
        0.20484488, 0.18685909, 0.25196077, 0.28350875, 0.344694  }}
    
    \foreach \i in {0,...,29} {
        \def \x {\axx+\dataaxsz*\xs[\i]}
        \def \y {\firstcolheight+\dataaxsz*4/5*\ys[\i]}
        \draw[thick, fill=black] (\x, \y) ellipse (1pt and 1pt);
    }
    \node[align=center] at (\axx + \dataaxsz/2 , \firstcolheight-.5) {Input-Output\\Data};

    \def \fcy{\firstcolheight + \ah/2}
    \def \offset{.35}

    \draw[->] (\ax+\firstcoloff+\tsw + \offset, \fcy) -- (\axx - \offset, \fcy);
    \draw[->] (\axx +\dataaxsz + \offset, \fcy) -- (\elx - \ah/2 - \offset, \fcy);
    \pgfmathsetmacro{\arrmid}{(\axx +\dataaxsz + \elx - \ah/2)/2}
    \draw[->] (\arrmid, \fcy+1) -- (\arrmid, \fcy+0.1);
    \node[align=center] at (\arrmid, \fcy+1.5) {Structural\\Constraints};    

    \draw[<-] (\aw+\pcroboff + 4*\ah/5 + \offset, \ely) -- (\elx - \uqeh *2 - .5 - \offset, \ely);
    \draw[<-] (2+\ax + \aw/2 + \offset, \ely) -- (\aw+\pcroboff - \offset, \ely);

    \draw[->] (\elx, \firstcolheight - \offset) -- (\elx, \ely + \uqeh + .5 +\offset) ;
    
\end{tikzpicture}
\caption{Illustration of the proposed D2PC framework.}
\label{fig:framework}
\end{figure*}
\section{Introduction}
MPC is widely used due to its inherent ability to handle constraints and its applicability to general multi-input multi-output systems. The key requirement for applying MPC is a model of the system, but obtaining such a model is often one of the most resource and labour intensive faces of the control design~\cite{ogunnaike1996contemporary}. This has led to a surge of interest within the research community on both direct~\cite{yin2023stochastic,breschi2023data, teutsch2026stochastic,berberich2020data, coulson2019data, coulson2021distributionally, Linbin2019, elokda2021data} and indirect~\cite{umenberger2019robust, terzi2019learning,lorenzen2019robust, koehler2022state, dorfler2023} data-driven control methods; i.e. strategies that primarily rely on data to design controllers. While data-driven methods offer significant benefits, their use in control raises challenges related to uncertainty quantification and constraint satisfaction~\cite{hou2013model}. Reliable controller synthesis requires both estimating system parameters from data and explicitly accounting for the induced uncertainty in robust and predictive control frameworks. We next review relevant work in the literature.

\textbf{Data-driven robust control: }
A crucial step in the development of data-driven controllers is robust control designs for the uncertain models obtained from data. 
Recent data-driven techniques utilize state measurements with energy bounded noise to synthesize robust state-feedback controllers without explicit system identification~\cite{van2020noisy}. 
This approach was further extended to incorporate known structural model constraints in~\cite{berberich2022combining}. 
However, these methods cannot deal with stochastic noise in the data. 
In contrast, \cite{umenberger2019robust} synthesizes robust state-feedback controllers using confidence sets derived through Bayesian regression. However, this uncertainty quantification and synthesis is limited to noise-free state measurements. 
In~\cite{barenthin2008identification}, the prediction error method is used to quantify parametric uncertainty from stochastic input-output data and a robust state-feedback controller is designed for a special class of parameterized systems.  
In contrast, the proposed approach synthesizes dynamic output-feedback controllers for a broad class of stochastic linear systems with partial measurements that robustly account for the uncertainty in the estimated model parameters.

\textbf{Data-driven predictive control}:
Indirect data-driven MPC techniques are well-established in the literature, however, results are typically  limited to bounded disturbances or noise-free state measurements~\cite{lorenzen2019robust,arcari2023stochastic,terzi2019learning}. 
In contrast, recent direct data-driven MPC methods (cf.~\cite{berberich2020data,yin2023stochastic,coulson2021distributionally,teutsch2026stochastic}) have gained traction as they enable direct prediction using input-output measurements. 
In~\cite{yin2023stochastic,coulson2021distributionally,teutsch2026stochastic}, (chance) constraints for finite-horizon open-loop problems are enforced. 
This is achieved using (implicit) multi-step predictors~\cite{koehler2022state}.  
Closed-loop guarantees are derived in~\cite{berberich2020data}, however, results are largely qualitative and conservative. 
In contrast, we propose an indirect data-driven predictive control framework that is applicable to input-output data with unbounded stochastic noise, exploits structured state-space models, and guarantees recursive feasibility, satisfaction of chance constraints, and an average expected cost bound for the resulting closed-loop system. 

\textit{Contribution:} The primary contribution of this work is D2PC, a framework that bridges data-driven techniques and predictive control through a design pipeline illustrated in Fig.~\ref{fig:framework}. Our approach is detailed in the following sections:
\begin{itemize}
    \item Section~\ref{sec:problem_setup} introduces the problem setup under consideration.
    \item Section~\ref{sec:sysid} presents our parameter identification method for stochastic linear systems with partial measurements that builds upon the Expectation-Maximization algorithm~\cite{GIBSON20051667,shumway-em}. The proposed method extends this line of work by integrating (general) structural constraints.
    \item Section~\ref{sec:unc_quant} outlines an asymptotic approximation to quantify uncertainty~\cite{newey1994large}, resulting in an uncertainty set over the estimated parameters.
    \item Section~\ref{sec:lqg} demonstrates our proposed method to design dynamic output-feedback controllers that is tailored for the established uncertainty set, leveraging the full-block S-procedure~\cite{scherer2000robust}. Additionally, we propose a simplified over-approximation of the uncertainty set that reduces computational complexity of the controller synthesis.
    \item Section~\ref{sec:pred_control} presents our predictive control scheme that ensures recursive feasibility and chance-constraint satisfaction. This framework, extends stochastic MPC methods~\cite{hewing2020recursively,muntwiler2023lqg,arcari2023stochastic} to jointly account for partial measurements and parametric uncertainties. Furthermore, an extensive theoretical analysis of the closed-loop properties of the proposed scheme is provided.
    \item Section~\ref{sec:example} presents a comprehensive walkthrough of the proposed framework, demonstrating its effectiveness through a numerical example involving a 10-dimensional spring-mass-damper system. We contrast our proposed framework to an established direct data-driven method~\cite{yin2023stochastic}.
    \item Section~\ref{sec:conclusion} concludes the paper.
\end{itemize}
Overall, the D2PC framework marks a significant advancement in integrating data-driven techniques with predictive control.  This work builds upon existing system identification and uncertainty quantification methods, developing tailored strategies to embed the resulting structured uncertainty into robust and predictive control synthesis.  In particular, a key contribution of our work is that the proposed robust control synthesis and predictive control method provide rigorous guarantees while remaining consistent with the setting and uncertainty quantification required in stochastic system identification methods. This ensures a principled integration of data-driven estimation techniques with model-based control design, preserving both robustness and consistency. A discussion of the related work corresponding to each section will be provided at the end of the respective sections. Alongside this paper, we provide a code framework that implements all the described steps for a general class of linear systems: \url{https://github.com/haldunbalim/D2PC}. 
\ifbool{arxiv}{}
{Additional details regarding the implementation of the identification scheme and the offline design can be found in an extended online version~\cite{balim2024data}.}

 
\textit{Notation:} We denote the set of real numbers as $\R$, natural numbers as $\N$, symmetric positive (semi-)definite matrices of size $n \times n$ as $\mathbf{S}_{++}^n$ ($\mathbf{S}_{+}^n$). Define $\mathrm{vec}(A) \in \mathbb{R}^{nm}$ as the operation that converts a matrix $A \in \mathbb{R}^{n \times m}$ into a vector by stacking its columns sequentially. Conversely, the operation $\mathrm{unvec}_{n}^{m}(x) \in \mathbb{R}^{m \times n}$ transforms a vector $x \in \mathbb{R}^{mn}$ back into a matrix by arranging every set of $m$ elements as columns of the resulting matrix. We use, $e_{n, i} \in \R^n$ to signify the $i$-th column of the identity matrix of dimension $n$. We denote the trace of a matrix $A$ by $\tr{A}$. For notational brevity, lower-triangular elements of symmetric matrices are denoted by $\star$. Additionally, for expressions involving symmetric forms $A^\top P A$, where $P \in \mathbb{R}^{n \times n}$ and $A \in \mathbb{R}^{n \times m}$, we use $[\star]^\top PA$ for notational convenience.  We denote the Moore-Penrose inverse of $ A $ as $ A^\dagger $ and use $ A \propto B $ to indicate direct proportion. For symmetric matrices, $Q \succeq 0$ ($Q \succ 0$) denotes positive semidefinite (positive definite). For $ Q \succeq 0 $, we write $ \|x\|_Q^2 = x^\top Q x $. The induced 2-norm  of a matrix $ A\in\mathbb{R}^{n\times m} $ is denoted by $ \|A\| $. A multivariate Gaussian vector $ x $ with mean $ \mu $ and covariance $ \Sigma $ is written as $ x \sim \mathcal{N}(\mu, \Sigma) $. We use $\Pr[X]$ for the probability of event $X$, $\E[X]$ for its expectation, and $\E[X \mid Y]$, $\Pr[X \mid Y]$ for conditional expectation and probability given $Y$. The identity matrix is denoted by $I$.

\section{Problem Setup} \label{sec:problem_setup}
We analyze uncertain discrete-time linear time-invariant (LTI) systems characterized by the following state-space representation:
\begin{align} \label{eq:state-space}
    x_{t+1} &= A(\vartheta)x_t + B(\vartheta)u_t + E w_t,\\
    y_t &= Cx_t + v_t,\notag\\
    w_t &\sim \mathcal{N}(0,Q(\eta)),\ v_t \sim \mathcal{N}(0, R(\eta)) ,\notag
\end{align}
with state $ x_t \in \mathbb{R}^{n_\mathrm{x}}$, control input $ u_t \in \mathbb{R}^{n_\mathrm{u}}$, measured output $ y_t \in \mathbb{R}^{n_\mathrm{y}}$, discrete-time index $t\in \N$, disturbance $ w_t\in \mathbb{R}^{n_\mathrm{w}}$, and measurement noise $v_t\in \mathbb{R}^{n_\mathrm{y}}$. The process and measurement noise vectors $w_t$, $v_t$ are assumed to be independent and identically Gaussian distributed with symmetric positive-definite covariance matrices. The matrix $C \in \R^{n_\mathrm{y} \times n_\mathrm{x}}$ and  $E \in \R^{n_\mathrm{x} \times n_\mathrm{w}}$ are assumed to be known and full rank. The system matrices $A(\vartheta)$ and $B(\vartheta)$ are affinely parameterized by the unknown vector $\vartheta$:
\begin{align}\label{eq:sys_mat_param}
    [A(\vartheta),\ B(\vartheta)] =[A_0,\ B_0]+E\mathrm{unvec}^{n_\mathrm{w}}_{n_\mathrm{x}+n_\mathrm{u}}(J\vartheta),
\end{align}
where $[A_0,\ B_0]$, $ J \in \mathbb{R}^{n_{\mathrm{w}}(n_{\mathrm{x}}+n_{\mathrm{u}})\times n_{\mathrm{\vartheta}}}$ are known matrices that define the parametrization of the system matrices by the unknown parameter vector $\vartheta \in\mathbb{R}^{n_{\mathrm{\vartheta}}}$. We note that, eq.~\eqref{eq:sys_mat_param} constrains  $\vartheta$ to parametrize the dynamics that are only in the span of disturbances ($Ew$); since, the state dimensions unaffected by disturbances can be estimated using few samples. We emphasize that if no structural information is available, the matrices can simply be chosen according to a canonical form~\cite{astrom1979maximum}, allowing us to model general LTI systems. The noise covariance matrices $Q(\eta)$ and $R(\eta)$ are parameterized by unknown vectors $\eta$. 

\begin{remark}[Model Generality and Special Cases]
    The parameterization~\eqref{eq:sys_mat_param} exemplifies a flexible approach for representing a wide class of LTI systems subject to various structural constraints. Our framework allows incorporation of known structural constraints when available and remains applicable in the absence of such prior knowledge. Notably, it encompasses two standard cases commonly considered in the literature:
    \begin{enumerate}
        \item Canonical Form: If there is no known structural information about the system matrices , they can be constrained to have a canonical form~\cite{levine1999control}, such as the (multi-variate) observable canonical form. 
        Such a parametrization is particularly useful if there is no known structural information about the system, except for the model order, and contains ARX models~\cite{box2015time} as an important special case.

        \item Structured Models: The considered setup accommodates a broad class of pre-defined constraints, such as affine constraints on system matrices. Additionally, the incorporation of the $E$ matrix enables handling semi-definite process noise covariance matrices, which allows certain dimensions of the process equations to be noise free. The flexibility of the considered parameterization will be further demonstrated in Section~\ref{sec:example}.
    \end{enumerate}
\end{remark}

The objective of this paper is to develop a comprehensive framework for data-driven control of the system described by~\eqref{eq:state-space}. The proposed framework includes the estimation of unknown parameters $\vartheta$ and $\eta$ with EM, quantification of uncertainties in the estimates, design of an output-feedback controller that robustly stabilizes the uncertain system, and formulation of a MPC scheme that guarantees chance constraint satisfaction while preserving the stability properties of the robust controller (cf. Fig~\ref{fig:framework}).

\section{Parameter Identification} \label{sec:sysid}

In this section, we provide a method to estimate the unknown parameter vector $\theta = (\vartheta, \eta)$ for the structured model~\eqref{eq:state-space} by adapting standard methods from stochastic estimation.  

We consider data generated from system~\eqref{eq:state-space} by applying a persistently exciting (cf.~\cite{coulson2022quantitative}) open-loop input sequence $u_t$ of length $T$. For the parameter identification, we utilize the resulting input-output trajectory $Y_T\coloneqq\{y_t\}_{t=1}^T$, $U_T\coloneqq\{u_t\}_{t=0}^{T-1}$. The initial state $x_0$ for this trajectory is assumed to follow a Gaussian distribution with unknown parameters, i.e., $x_0 \sim \mathcal{N}(\bar{x}_0(\eta), \Sigma_{\mathrm{x},0}(\eta))$ with $\Sigma_{\mathrm{x},0}(\eta) \in \mathbf{S}_{++}^{n_\mathrm{x}}$.

\textit{Maximum Likelihood Estimation (MLE)} is a well-established method for parameter estimation, which is typically asymptotically optimal, achieving the Cramér-Rao bound~\cite{Ljung1998,cramer1999mathematical}. The MLE is formally defined as:
\begin{equation}\label{eq:mle}
    \hat\theta_{\mathrm{MLE}} = \argmax_{\theta \in \Theta} p_\theta(Y_\mathrm{T})
\end{equation}
where $\Theta$ is set of considered parameter vectors, and $p_\theta(Y_\mathrm{T})$ denotes the likelihood of the given output trajectory evaluated with the parameters $\theta$. We assume that the true system parameters satisfy $\theta\in \Theta$. The covariance matrices $Q(\eta)$, $R(\eta)$, and $\Sigma_{\mathrm{x},0}(\eta)$ are positive-definite $\forall \theta \in \Theta$. The MLE problem~\eqref{eq:mle} is a non-convex optimization problem due to the concurrent estimation of states and parameters.


\textit{Expectation-Maximization (EM):} In the following, we briefly outline the EM algorithm, adapting~\cite{GIBSON20051667} to account for structural constraints (cf. (Sec.~\ref{sec:problem_setup})). 
We are searching for the parameters that maximizes the likelihood for the given measurement trajectory $Y_{\mathrm{T}}$. Denote the log-likelihood of the measurement trajectory using the parameters $\theta$ as $\log p_{\mathrm{\theta}}(Y_T)$. Furthermore, define $X_T = \{x_t\}_{i=0}^{T}$ to be the corresponding state trajectory. Respectively, given a parameter vector $\theta$, the associated likelihood can be equivalently stated based on the expected value conditioned on $\theta^\prime$, with some arbitrary parameter vector $\theta^\prime$:
\begin{align}
    L(\theta) =& \E[\log p_\theta(Y_\mathrm{T})\mid\theta^\prime, Y_\mathrm{T}] = \E[\log p_\theta(X_\mathrm{T}, Y_\mathrm{T}) - \log p_\theta(X_\mathrm{T}\mid Y_\mathrm{T})\mid\theta^\prime, Y_\mathrm{T}], \notag
\end{align}
where the expectation is taken over the realizations of the process noise, measurement noise, and initial state distribution. 
Consequently, the difference of log-likelihood for two different parameters $\theta$, $\theta^\prime$ can be equivalently written as:
\begin{equation}\label{eq:diff_ll}
    L(\theta) - L(\theta^\prime) = \mathcal{Q}(\theta, \theta^\prime) - \mathcal{Q}(\theta^\prime, \theta^\prime) + \mathrm{KL}(p_\theta \lvert\rvert p_{\theta^\prime}),
\end{equation}
where $\mathcal{Q}(\theta, \theta^\prime)$ denotes the conditional log-likelihood and $\mathrm{KL}(p_\theta \lvert\rvert p_{\theta^\prime})$ denotes the Kullback-Leibler divergence~\cite{mackay2003information}, which are defined as:
\begin{align}\label{eq:cond_ll_defn}
    \mathcal{Q}(\theta, \theta^\prime) &= \E[\log p_\theta(X_\mathrm{T}, Y_\mathrm{T}) \mid \theta^\prime, Y_\mathrm{T}], \quad
    \mathrm{KL}(p_\theta \lvert\rvert p_{\theta^\prime}) = \E\left[\log \left(\frac{p_{\theta^\prime}(X_\mathrm{T}\mid Y_\mathrm{T})}{p_\theta(X_\mathrm{T} | Y_\mathrm{T})}\right) \ \middle |\ \theta^\prime, Y_\mathrm{T}\right].
\end{align}
Using the Kullback-Leibler divergence's non-negativity property, it holds that:
\begin{equation}\label{eq:em-main}
    L(\theta) - L(\theta^\prime) \geq \mathcal{Q}(\theta, \theta^\prime) - \mathcal{Q}(\theta^\prime, \theta^\prime).
\end{equation}
From equation~\eqref{eq:em-main}, it is apparent that increasing the conditional log-likelihood function $\mathcal{Q}(\theta, \theta^\prime)$ also increases the likelihood. Based on this principle, the Generalized EM (GEM) algorithm is summarized in Algorithm~\ref{algo:em}.
\begin{algorithm}
\caption{Generalized EM Algorithm}
\label{algo:em}
\begin{algorithmic}[1]
 \State \textbf{Input:} stop tolerance $\epsilon \geq 0$, initial estimate $\theta_0\in\Theta$
\While{$L(\theta_{k}) - L(\theta_{k-1}) \geq \epsilon$}
    \Statex \% Kalman Smoother
    \State \textit{E-Step:} Construct $\mathcal{Q}(\theta, \theta_k)$. 
    \Statex \% Analytical solution or iterative optimization
    \State \textit{GM-Step:} Compute $\theta_{k+1} = \mathtt{GM}(\theta_k)$.\label{algo_line:gm}
\EndWhile
\end{algorithmic}
\end{algorithm}\\
The original EM algorithm, directly computes the maximizer to the surrogate function $\mathcal{Q}(\theta, \theta_k)$. However, depending on the structural constraints, it is not always possible to analytically compute the unique global maximizer. The Generalized M-step (GM) addresses this issue by applying an algorithm guaranteeing a monotonic increase in the conditional log-likelihood at each iteration~\cite{dempster1977maximum}. In particular, the maximization is replaced by any algorithm $\mathtt{GM}:\Theta\rightarrow\Theta$ with the following property:
\begin{equation}\label{eq:cond_oracle_alg}
    \mathcal{Q}(\mathtt{GM}(\theta_k), \theta_k) \geq \mathcal{Q}(\theta_k, \theta_k), \  \forall \theta_k \in \Theta,
\end{equation}
where the condition holds with equality if and only if $\theta_k$ is a local minima of $\mathcal{Q}(\theta, \theta_k)$ over $\Theta$. The E and GM steps follow a standard procedure.   We provide an efficient implementation within our open-source code\ifbool{arxiv}{ (see also~App.~\ref{sec:gem_deets})}{}.
\begin{proposition}[{Adapted from~\cite[Theorem 1]{wu1983convergence}}]\label{prop:em_convergence}
 Consider the parameter sequence generated by Algorithm~\ref{algo:em} with $\mathtt{GM}$ satisfying~\eqref{eq:cond_oracle_alg}. Then, the likelihood, $L(\theta_k)$ increases monotonically. Furthermore, if $\Theta$ is compact, and $\epsilon = 0$, Algorithm~\ref{algo:em} converges to a stationary point of the log-likelihood function.
\end{proposition}

\textit{Discussion:} The literature offers a diverse array of methods to tackle the MLE problem~\cite{astrom1979maximum}. For instance, sampling-based approaches like particle filters and Markov Chain Monte Carlo based methods are explored in~\cite{mcmc-mle2} and~\cite{ninness2010bayesian}. However, these approaches require large number of samples to accurately model the likelihood function, especially for high-dimensional problems. In contrast, the EM algorithm, discussed in~\cite{GIBSON20051667,shumway-em}, scales to high-dimensional problems with moderate computational complexity.

Another widely utilized approach for MLE is the Prediction Error Method~\cite{astrom1979maximum, simpson2022parameter}. This technique directly optimizes the likelihood using nonlinear programming.\footnote{%
Recent results suggest that local optimization procedures may be sufficient to ensure asymptotic consistency of the estimated observer gain ~\cite{simpson2025identification}.} However, a primary limitation of these methods is their computational cost as data size increases. Conversely, the EM algorithm is less affected by increasing data sizes, since the conditional log-likelihood function $\mathcal{Q}(\theta, \theta^\prime)$ is independent of the data size. Consequently, the computational complexity of a single EM iteration scales linearly with respect to data size $T$.

Another strategy to estimate dynamical models is using a large enough set of past input-output measurements to represent the internal state~\cite{Ljung1998}. 
This reduces the estimation to a least-squares problem and the computational efficiency and simplicity has motivated much recent work on direct data-driven methods with this parametrization~\cite{coulson2021distributionally, breschi2023data,chiuso2023harnessing}. However, such approaches do not allow for the incorporation of the structural constraints~\eqref{eq:sys_mat_param} and the resulting high state dimension would yield scalability issues in the later control design.


\section{Uncertainty Quantification}\label{sec:unc_quant}
To design reliable controllers, we need to determine a set $\Theta_\delta$ containing the uncertain parameters $\vartheta$ with a user-chosen probability level $\delta$. 
In the following, we describe a standard strategy to quantify the uncertainty over the estimated parameters. 
We compute $\hat{\vartheta}$ using the EM algorithm, which monotonically increases the likelihood and, when initialized sufficiently close to the global optimum, converges to the MLE. 
Consistency of the MLE for LTI models can be ensured under standard assumptions (see, e.g., \cite{Ljung1998}), which include among others identifiability of the model\footnote{%
Additional standard requirements include: (i) persistence of excitation of the input, i.e., $\sum_{k=t}^{t+T-1} u_k u_k^\top \succeq cI$ for some $c, T > 0$ and all $t$; (ii) independence of the input from the process and measurement noise; and (iii) the existence of a parameter vector $\vartheta$ such that the data-generating system coincides with the parametrization~\eqref{eq:state-space}, including the noise covariances.
}. 
In the following, we assume that the estimate $\hat{\vartheta}$ is consistent and  we leverage the asymptotic properties of consistent estimators.

For consistent estimates, the asymptotic distribution of the deviation $\tilde \vartheta\coloneqq \vartheta-\hat\vartheta$ is Gaussian with zero mean and its covariance is defined by the inverse of the expected Fisher information matrix~\cite{vaart_1998}:
\begin{equation}\label{eq:exp_fish_true}
    H(\vartheta) = -\E\left[\frac{\partial^2}{\partial \vartheta^2} \log p_{\vartheta}(Y_T)\right],
\end{equation}
and the matrix  $H(\vartheta)$ is positive-definite, assuming identifiability.  
Since $\vartheta$ is not known, we approximate the expected Fisher information matrix with the observed information matrix $\hat H(\hat{\vartheta})$ evaluated at ML estimate $\hat{\vartheta}$, as suggested by~\cite{newey1994large}:
\begin{equation}\label{eq:obs_fish_hat}
    \hat H(\hat \vartheta) = -\left.\frac{\partial^2}{\partial \vartheta^2} \log p_{\vartheta}(Y_T)\right|_{\vartheta=\hat\vartheta}.
\end{equation}
Similarly to $H(\vartheta)$, we assume that $\hat H(\hat \vartheta)$ is strictly positive-definite. Accordingly, we approximate the uncertainty as $\vartheta \sim \mathcal{N}(\hat \vartheta, \hat H^{-1}(\hat \vartheta))$. If $\hat{\vartheta}$ is a consistent estimator and $H(\vartheta)$ is continuous, then the derived distribution for the parameters $\vartheta$ is  asymptotically correct~\cite{newey1994large}. 
Under this approximation, the true system parameters lie within an ellipsoidal confidence region around $\hat{\vartheta}$ with probability $\delta \in (0, 1)$, i.e., $\Pr[\vartheta \in \Theta_{\delta}] \geq \delta$ with
\begin{equation} \label{eq:param_set}
    \Theta_{\delta} = \{\vartheta \mid (\vartheta - \hat \vartheta)^\top \Sigma_{\vartheta, \delta}^{-1} ( \vartheta - \hat \vartheta) \leq 1\},
\end{equation}
where $\Sigma_{\vartheta, \delta} \coloneqq \chi^2_{n_{\vartheta}}(\delta)\Sigma_{\vartheta}$, $\Sigma_{\vartheta} = H^{-1}(\hat{\vartheta}) \succ 0$, and $\chi^2_{n_{\vartheta}}$ denotes the quantile function of the chi-squared distribution with $n_{\vartheta}$ degrees of freedom. 
For the following design, we assume that these approximations are provided as a valid description of the uncertainty, which is in general only true under additional identifiability conditions and in the asymptotic limit of large data. 
This assumption is formalized below. 
\begin{assumption}\label{asm:param_set}
    The covariance matrices for measurement and process noise are known or over-estimated; i.e. $Q(\hat \eta) \succeq Q(\eta)$, $R(\hat \eta) \succeq R(\eta)$. The true parameters $\vartheta$ are confined within a known ellipsoidal set $\Theta_\delta$ from~\eqref{eq:param_set}.
\end{assumption}

Asm.~\ref{asm:param_set}, establishes a set over the unknown vector $\vartheta$. For the remainder of the paper we suppose that Asm.~\ref{asm:param_set} holds. 
For this work, we do not consider the uncertainty in the variance estimate and we denote $Q = Q(\hat \eta)$, $R=R(\hat \eta)$. To synthesize a robust controller we establish a parametric uncertainty set. For this purpose, we approximately quantify uncertainty by modeling it as a Gaussian distribution. This allows the parameters to be contained within a user-defined probability level $\delta$; consequently, we expect $\vartheta \in \Theta_\delta$ to hold with probability approximately $\delta$.

\textit{Discussion:} Note that the proposed uncertainty characterization is an approximation, assuming identifiability of the model and asymptotic limit. Nonetheless, confidence bounds based on this approximation often provide a reasonable approximation, especially when the estimated parameters are close to their true values. The reliability of this approximation will later be empirically evaluated in a numerical example in Sec.~\ref{sec:example}. The outlined strategy has also been used to derive uncertainty over the parameter estimates with EM algorithm~\cite{efron1978assessing}. Furthermore, in~\cite{holmes2013derivation} this strategy has been adopted for uncertainty characterization for the parameters of the state-space models. In the special case of state measurement ($y=x$), the parameter estimation problem can be reduced to linear regression and strong finite-data error bounds are available, see for example~\cite{abbasi2011improved}. 


\section{Robust Controller Synthesis}\label{sec:lqg}
In this section, we first derive a linear fractional representation for the system described by~\eqref{eq:state-space}, taking into account the parameter set specified in Asm.~\ref{asm:param_set}. Subsequently, we present a methodology for synthesizing a robust dynamic output-feedback controller.

\subsection{Linear Fractional Representation}

In this subsection, we will construct a linear fractional representation~\cite{zhou1998essentials} for the uncertain system~\eqref{eq:state-space}. The following lemma establishes the relation between system matrices and $\vartheta$.
\begin{lemma}\label{lem:ab_param}
    The system matrices satisfy
    \begin{equation}\label{eq:ab_defn}
        [A(\vartheta),\ B(\vartheta)] = [\hat A,\ \hat B] + E\Delta J_\Delta,
    \end{equation}
    with $\Delta = I_{n_\mathrm{w}} \otimes \tilde \vartheta^\top$, $\tilde \vartheta =  \vartheta - \hat \vartheta$, $[\hat A, \hat B]$ denoting system matrices associated with mean parameter estimate $\hat \theta$, and
    \begin{equation}\label{eq:Jdelta}
        J_\Delta\coloneqq(I_{n_\mathrm{w}}\otimes (P_{n_\mathrm{x}+n_\mathrm{u}}^{ n_\mathrm{w}}J)^\top) (\mathrm{vec}(I_{n_\mathrm{w}}) \otimes I_{n_\mathrm{x}+n_\mathrm{u}}),
    \end{equation}
    where $P^{m}_n \in \R^{mn \times mn}$ denotes the commutation matrix.
\end{lemma}
\begin{proof}
From~\eqref{eq:sys_mat_param} and the definition of $[\hat A,\hat B]$, we obtain
\begin{equation}
[A(\vartheta),\,B(\vartheta)]
= [\hat A,\,\hat B]
+ E\,\unvec^{n_\mathrm{w}}_{n_\mathrm{x}+n_\mathrm{u}}(J\tilde\vartheta).
\end{equation}
Hence, it remains to show that $\unvec^{n_\mathrm{w}}_{n_\mathrm{x}+n_\mathrm{u}}(J\tilde\vartheta) = \Delta J_\Delta$.
This identity follows directly from the standard commutation and reshaping property of the $\unvec$ operator; see Lemma~\ref{lem:U_defn} for details.
\end{proof}
Thus, we can represent the system~\eqref{eq:state-space} using the following linear fractional representation:
\begin{align}\label{eq:open_loop_lfr}
    \begin{bmatrix}
        x_{t+1} \\ y_t \\ q_t
    \end{bmatrix} &= 
    \begin{bmatrix}
        \hat A & \hat B & E & E & 0 \\ C & 0 & 0 & 0 & I \\ \multicolumn{2}{c}{J_\Delta}  & 0 & 0 & 0
    \end{bmatrix}
    \begin{bmatrix}
        x_t \\ u_t \\ p_t \\ w_t \\ v_t
    \end{bmatrix},\ p_t = \Delta q_t 
\end{align}
where $p_t$ represents the effect of parametric uncertainty~\cite{scherer2000robust}.  
In~\cite[Prop.2]{strasser2023control}, a multiplier set for Kronecker products is developed. Inspired by this, the following lemma establishes an equivalent uncertainty set over the matrices $\Delta$.
\begin{lemma}\label{lem:multiplier}
    Consider the set
\begin{align}\label{eq:param_set_delta}
    \boldsymbol{\Delta}_\delta =\Biggl\{\Delta \in \R^{n_\mathrm{w} \times n_\mathrm{w}n_\mathrm{\vartheta}} \:\Bigg|\: &\begin{bmatrix}
        \Delta^\top \\I_{n_\mathrm{w}} 
    \end{bmatrix}^\top P_{\Delta, \delta} \begin{bmatrix}
        \Delta^\top \\ I_{n_\mathrm{w}} 
    \end{bmatrix}\succeq 0,  ~ \forall P_{\Delta, \delta} \in \mathbf{P}_{\Delta, \delta} \Biggr\}
\end{align}
with the multipliers set:
\begin{align}\label{eq:PDelta}
    \mathbf{P}_{\Delta, \delta} = \left\{\begin{bmatrix}
       -\Lambda \otimes \Sigma_{\vartheta,\delta}^{-1} & 0 \\ 0 &  \Lambda
    \end{bmatrix} \:\middle |\: 0 \preceq \Lambda \in \R^{n_\mathrm{w}\times n_\mathrm{w}} \right\}.
\end{align}
    Then, $\Delta \in \boldsymbol{\Delta}_\delta$ if and only if $\Delta =I_{n_\mathrm{w}}\otimes \tilde \vartheta^\top$ with $\vartheta \in \Theta_\delta$.
\end{lemma}
\begin{proof}
\textit{"If":} Suppose that $\vartheta \in \Theta_\delta$ and let $\Lambda \succeq 0$ be arbitrary, then $I_{n_\mathrm{w}}\otimes \tilde \vartheta^\top$ satisfies:
\begin{align}
    &\begin{bmatrix}
            I_{n_\mathrm{w}}\otimes \tilde \vartheta  \\ I_{n_\mathrm{w}} 
        \end{bmatrix}^\top 
          \begin{bmatrix}
        -\Lambda \otimes \Sigma_{\vartheta,\delta}^{-1} & 0 \\ 0 & \Lambda
    \end{bmatrix} \begin{bmatrix}
            I_{n_\mathrm{w}}\otimes \tilde \vartheta \\ I_{n_\mathrm{w}}
        \end{bmatrix}
     = \Lambda - (I_{n_\mathrm{w}}\otimes \tilde \vartheta^\top) (\Lambda \otimes \Sigma_{\vartheta,\delta}^{-1})(I_{n_\mathrm{w}}\otimes \tilde \vartheta) \notag\\
     &= \Lambda - \Lambda \otimes(\tilde \vartheta^\top \Sigma_{\vartheta,\delta}^{-1} \tilde \vartheta) \notag
     = \Lambda (1-\tilde \vartheta^\top \Sigma_{\vartheta,\delta}^{-1} \tilde \vartheta) \succeq 0 \notag, \quad
     \iff (\vartheta -\hat \vartheta)^\top \Sigma_{\vartheta,\delta}^{-1} (\vartheta -\hat \vartheta)\leq 1 .
\end{align}
The final step invokes that condition holds $\forall\Lambda \succeq 0$. This establishes that $I_{n_{\mathrm{w}}} \otimes \tilde{\vartheta}^\top \in \boldsymbol{\Delta}_\delta$ for all $\vartheta \in \Theta_\delta$. The only if case follows from~\cite[Prop. 2]{strasser2023control}. 
\end{proof}
By defining the set $\boldsymbol{\Delta}_\delta$ to have a bijective correspondence with the original set $\Theta_\delta$, 
we provide an equivalent representation of the parametric uncertainty resulting from the estimation (Asm.~\ref{asm:param_set}), making it suitable for application in robust control techniques.
\subsection{Robust Output-Feedback Controller Synthesis}
\begin{figure}[t]
\begin{center}
    \begin{tikzpicture}

\def\vsep{.25};
\def\osz{1.5};
\def\syssz{1.6};
\def\arrowh{.6};
\def\txty{.25};
\def\perc{.2};
\def\ctrlwo{.25};

\pgfmathsetmacro{\x}{(\syssz -\osz)/2+\arrowh}
\pgfmathsetmacro{\xc}{\x-\ctrlwo/2}
\pgfmathsetmacro{\y}{\vsep*2+\osz+\syssz}
\pgfmathsetmacro{\lowarr}{\osz+\vsep+\syssz*\perc}
\pgfmathsetmacro{\uparr}{\osz+\vsep+\syssz-\syssz*\perc}

\pgfmathsetmacro{\txtx}{\arrowh/2}

\draw[thick, rounded corners] (\xc, \y) rectangle ++(\osz+\ctrlwo, \osz) node[pos=.5, align=center]{Controller};
\draw[thick, rounded corners] (\x,0) rectangle ++(\osz, \osz) node[pos=.5, align=center]{$\Delta$};

\draw[thick, rounded corners] (\arrowh,\vsep +\osz) rectangle ++(\syssz, \syssz) node[pos=.5, align=center]{System};
\draw[thick,-] (\arrowh,\lowarr) -- (0,\lowarr);
\draw[thick,-] (0,\lowarr) -- (0,\osz/2);
\draw[thick,->] (0,\osz/2) -- (\x,\osz/2);

\draw[thick,->] (\arrowh*2+\syssz,\lowarr) -- (\arrowh+\syssz,\lowarr);
\draw[thick,-] (\arrowh*2+\syssz,\lowarr) -- (\arrowh*2+\syssz,\osz/2);
\draw[thick,-] (\arrowh*2+\syssz,\osz/2) -- (\x+\osz,\osz/2);

\draw[thick,-] (\arrowh,\uparr) -- (0,\uparr);
\draw[thick,-] (0,\uparr) -- (0,\y+\osz/2);
\draw[thick,->] (0,\y+\osz/2) -- (\xc,\y+\osz/2);

\draw[thick,->] (\arrowh*2+\syssz,\uparr) -- (\arrowh+\syssz,\uparr);
\draw[thick,-] (\arrowh*2+\syssz,\uparr) -- (\arrowh*2+\syssz,\y+\osz/2);
\draw[thick,-] (\arrowh*2+\syssz,\y+\osz/2) -- (\xc+\osz+\ctrlwo,\y+\osz/2);

\draw[thick,->] (\arrowh,\syssz/2+\vsep+\osz) -- (0,\syssz/2+\vsep+\osz);
\draw[thick,->] (\arrowh*2+\syssz,\syssz/2+\vsep+\osz) -- (\arrowh+\syssz,\syssz/2+\vsep+\osz);

\node[align=center] at (\txtx, \txty+\uparr) {$y$};
\node[align=center] at (\txtx, \txty+\lowarr) {$q$};
\node[align=center] at (\txtx+\syssz+\arrowh, \txty+\uparr) {$u$};
\node[align=center] at (\txtx+\syssz+\arrowh, \txty+\lowarr) {$p$};
\node[align=center] at (\txtx, \txty+\syssz/2+\vsep+\osz) {$\epsilon$};
\node[align=center] at (\txtx+\syssz+\arrowh, \txty+\syssz/2+\vsep+\osz) {$d$};

\end{tikzpicture}
\end{center}
\caption{Illustration of the linear fractional representation for the closed loop system~\eqref{eq:closed_loop_lfr}.} 
\label{fig:LFT}
\end{figure}
In this section, we design a robust dynamic output-feedback controller of the form:
\begin{align} \label{eq:control_law}
    x_{t+1}^{\mathrm{c}} &=A_\mathrm{c} x_t^{\mathrm{c}} + L y_t, \quad
    u_t = K x^c_{t} 
\end{align}
with controller state $x_{t}^{\mathrm{c}}\in\R^{n_\mathrm{x}}$ and design parameters $A_\mathrm{c}$, $K$, $L$. The goal is to design a controller that robustly stabilizes the system~\eqref{eq:open_loop_lfr} and minimizes the $\mathcal{H}_2$-norm of channel $d\rightarrow\epsilon$, with the performance output:
\begin{equation}\label{eq:perf_out}
    \epsilon_t = C_\mathrm{\epsilon}x_t+D_\mathrm{\epsilon}u_t\in\R^{n_\epsilon}. 
\end{equation}
 We can represent the closed-loop dynamics of the system~\eqref{eq:open_loop_lfr} using the following linear fractional representation:
\begin{align}\label{eq:closed_loop_lfr}
    \begin{bmatrix}
        \xi_{t+1} \\ \epsilon_t \\ q_t
    \end{bmatrix} &= \begin{bmatrix}
        \mathcal{\hat A} && \mathcal{B}_\mathrm{p} && \mathcal{B}_\mathrm{d} \\
        \mathcal{C}_\epsilon && 0 && 0 \\
        \mathcal{C}_\mathrm{q} && 0 && 0
    \end{bmatrix}\begin{bmatrix}
        \xi_t \\ p_t \\ d_t
    \end{bmatrix},\ p_t = \Delta q_t 
\end{align}
with:
\begin{align}\label{eq:clc_matrices}
    &\mathcal{\hat A} = \begin{bmatrix}
        \hat A & \hat BK \\ LC & A_\mathrm{c}
    \end{bmatrix}, \ \mathcal{B}_\mathrm{d} = \begin{bmatrix}
        EQ^{1/2} & 0 \\ 0 & LR^{1/2}
    \end{bmatrix},\  
    \mathcal{B}_\mathrm{p} = \begin{bmatrix}
        E \\ 0 
    \end{bmatrix},\ \mathcal{C}_{\mathrm{q}} = J_{\Delta} \begin{bmatrix}
        I& 0 \\ 0& K
    \end{bmatrix}, \ \mathcal{C}_{\mathrm{\epsilon}} =\begin{bmatrix}
       C_\mathrm{\epsilon} & D_\mathrm{\epsilon} K
    \end{bmatrix}, \
    \xi_t = \begin{bmatrix}
        x_t \\ x_t^{\mathrm{c}}
    \end{bmatrix},\ d_t = \begin{bmatrix}
        d^\mathrm{w}_t \\ d^\mathrm{v}_t
    \end{bmatrix},\ d_t \sim \mathcal{N}(0,I). 
\end{align}
This interconnection is also visualized in Figure~\ref{fig:LFT}. 
The following result ensures an upper bounds on the $\mathcal{H}_2$-norm based on the uncertainty parameterization in~\eqref{eq:param_set_delta}.
\begin{theorem}\label{thm:feas_control}
    Suppose that there exists $\mathcal{X} \in \mathbf{S}^{2n_\mathrm{x}}_{++}$, $\Lambda\in \mathbf{S}^{n_\mathrm{w}}_{++}$, $\gamma>0$ such that:
\begin{subequations}
\begin{equation}\label{eq:perf_cond_first} 
      \tr{\mathcal{C}_\epsilon \mathcal{X} \mathcal{C}_\epsilon^\top} \leq \gamma^2,
\end{equation}
    \begin{equation}\label{eq:stab_cond_nonlin}
    \begingroup 
        \setlength\arraycolsep{.65 pt}
       \begin{bmatrix}  \\  \\ \star \\  \\ \\
        \end{bmatrix}^\top
        \left[\begin{array}{cc|cc}
        \mathcal{B}_\mathrm{d}\mathcal{B}_\mathrm{d}^\top - \mathcal{X} & 0 & 0 & 0\\
        0 & \mathcal{X} & 0 & 0 \\
        \hline
        0 & 0 & -\Lambda \otimes \Sigma_{\vartheta,\delta}^{-1}  & 0 \\
        0 & 0 & 0 & \mathcal{B}_\mathrm{p}\Lambda\mathcal{B}_\mathrm{p}^\top
        \end{array}\right]
        \begin{bmatrix} I & 0  \\
            \mathcal{\hat A}^\top & \mathcal{C}_\mathrm{q}^\top \\
            \hline
            0 & I  \\
            I & 0 
        \end{bmatrix}\endgroup \prec 0.
    \end{equation}
\end{subequations}
Then, the closed-loop system described by equation~\eqref{eq:closed_loop_lfr} is Schur stable and the $\mathcal{H}_2$-norm of $d\rightarrow \epsilon$ is bounded by $\gamma$ for all $ \Delta \in \boldsymbol{\Delta}_\delta$.
\end{theorem}
\begin{proof}
Following~\cite[Lemma 1]{berberich2022combining}, $\mathcal{B}_\mathrm{p}$ being full column rank ensures that set $\boldsymbol{\tilde \Delta}_{\delta} = \mathcal{B}_\mathrm{p}\boldsymbol{\Delta}_{\delta}$ can also be characterized with multipliers $\Lambda$ similar to~\eqref{eq:PDelta}, see Lemma~\ref{lem:Tdelta} for details. 
The full-block S-procedure~\cite{scherer2000robust} ensures that~\eqref{eq:stab_cond_nonlin}  with $\Lambda\in\mathbf{S}_{++}^{n_\mathrm{w}}$ implies:
    \begin{align}\label{eq:lpv_delta}
        \begin{bmatrix}
            I \\  \tilde \Delta^\top
        \end{bmatrix}^\top \left(\begin{bmatrix}
            \mathcal{A} \\ \mathcal{C}_\mathrm{q}
        \end{bmatrix}\mathcal{X} \begin{bmatrix}
            \mathcal{A} \\ \mathcal{C}_\mathrm{q}
        \end{bmatrix}^\top +\begin{bmatrix}
            \mathcal{B}_\mathrm{d}\mathcal{B}_\mathrm{d}^\top-\mathcal{X}  & 0 \\ 0 & 0
        \end{bmatrix}\right) &
        \begin{bmatrix}
            I \\ \tilde \Delta^\top
        \end{bmatrix} \prec 0,\quad        \forall \tilde \Delta \in \boldsymbol{\tilde\Delta}_{\delta}&.
    \end{align}
    Then equation~\eqref{eq:lpv_delta} is equivalent to $\forall \Delta \in \boldsymbol{\Delta}_\delta$:
    \begin{equation}\label{eq:lpv}
        \mathcal{A}(\Delta) \mathcal{X} \mathcal{A}(\Delta)^\top - \mathcal{X} + \mathcal{B}_\mathrm{d}\mathcal{B}_\mathrm{d}^\top \prec 0
    \end{equation}
    where:
    \begin{equation}\label{eq:tilde_Adelta}
        \mathcal{A}(\Delta) = \mathcal{\hat A} +  \mathcal{B}_\mathrm{p}\Delta\mathcal{C}_\mathrm{q}. 
    \end{equation}
     Equation~\eqref{eq:lpv} shows that there exists a common Lyapunov function for the closed-loop system~\eqref{eq:closed_loop_lfr} $\forall \Delta \in \boldsymbol{\Delta_\delta}$~\cite[Thm. 10.1]{scherer2000robust}; thus, the Schur stability of $\mathcal{A}(\Delta)$ is guaranteed. Furthermore, together with~\eqref{eq:lpv}, the condition in~\eqref{eq:perf_cond_first} ensures the $\mathcal{H}_2$-norm for the channel $d \rightarrow \epsilon$ is bounded by $\gamma$, $\forall \Delta \in \boldsymbol{\Delta_\delta}$, see~\cite[Thm. 10.3]{scherer2000robust}.
\end{proof}
The provided theorem and proof closely follow~\cite[Thm. 1]{berberich2022combining} and extend it to dynamic output-feedback controllers with common procedures from the literature~\cite{scherer2000robust}. Our main contribution is to incorporate the multiplier set~\eqref{eq:PDelta}, thus extending the standard tools from robust control to the uncertainty set $\Theta_\delta$ resulting from the identification. We note  feasibility of \eqref{eq:stab_cond_nonlin} necessitates that the systems within the set $\boldsymbol{\Delta_\delta}$ are jointly stabilizable. Given Assumption~\ref{asm:param_set}, Theorem~\ref{thm:feas_control} provides a bound on the $\mathcal{H}_2$-norm of the true system in closed loop with the controller~\eqref{eq:control_law}. 

The synthesis of output-feedback controllers for systems with parametric uncertainties has been thoroughly investigated in the literature~\cite{scherer2000robust, zhou1998essentials} and has been recognized as a non-convex optimization problem. The controller design is facilitated through a process of alternating between robust synthesis and analysis, see \ifbool{arxiv}{Appendix~\ref{sec:lqg_synth}}{\cite[Appendix~C]{balim2024data}} for implementation details. 
To alleviate computational burden, the following proposition provides an over-approximation of the set $\boldsymbol{\Delta}_\delta $ that reduces the optimization problem's dimensionality.
\begin{proposition}\label{prop:alternative_unc_set}
    For any matrix $D \in \mathbf{S}_{++}^{n_\mathrm{x} + n_\mathrm{u}}$, consider the following set:
    \begin{equation}\label{eq:bar_Delta}
        \boldsymbol{\bar\Delta}_\delta = \left\{\bar \Delta \in \R^{n_\mathrm{w}\times (n_\mathrm{x}+n_\mathrm{u})} \mid \bar \Delta  D \bar\Delta^\top \preceq \lambda_{\mathrm{max}}(M) I\right\}
    \end{equation}
    with:
    \begin{equation}\label{eq:M_defn}
    M = \Sigma_{\vartheta, \delta}^{1/2} J^\top (D \otimes I) J \Sigma_{\vartheta, \delta}^{1/2}.
\end{equation}
    Then, $\boldsymbol{\Delta}_\delta J_\Delta \subseteq \boldsymbol{\bar\Delta}_\delta$, with $J_\Delta$ as in~\eqref{eq:Jdelta}.
\end{proposition}
\begin{proof}
Consider an arbitrary $\Delta \in \boldsymbol{\Delta_\delta}$, then the following relationships hold:
\begin{align}
    &\tr{\Delta J_\Delta  D J_\Delta^\top \Delta^\top}
    = \sum_{i=1}^{n_\mathrm{w}} e_{n_{\mathrm{w}},i}^\top \Delta J_\Delta DJ_\Delta^\top  \Delta^\top e_{n_{\mathrm{w}},i}
    = \sum_{i=1}^{n_\mathrm{w}} \mathrm{vec}(\Delta J_\Delta)^\top (I \otimes e_{n_{\mathrm{w}},i}) D (I \otimes e_{n_{\mathrm{w}},i})^\top \mathrm{vec}(\Delta J_\Delta)  \\
    \stackrel{\text{Lem.~\ref{lem:ab_param}}}{=}& \sum_{i=1}^{n_\mathrm{w}} \tilde{\vartheta}^\top J^\top (I \otimes e_{n_{\mathrm{w}},i}) D (I \otimes e_{n_{\mathrm{w}},i})^\top J \tilde{\vartheta}
    = \sum_{i=1}^{n_\mathrm{w}} \tilde{\vartheta}^\top J^\top (D \otimes e_{n_{\mathrm{w}},i} e_{n_{\mathrm{w}},i}^\top) J \tilde{\vartheta} \notag
    = \tilde{\vartheta}^\top J^\top (D \otimes I) J \tilde{\vartheta} 
    \leq \max_{\vartheta \in \Theta_\delta}  \tilde{\vartheta}^\top J^\top (D \otimes I) J \tilde{\vartheta} = \lambda_{\mathrm{max}} (M). 
\end{align}
Given that the trace of a symmetric positive semi-definite matrix is an upper bound to its maximum eigenvalue, we can deduce that $\Delta J_\Delta \in \boldsymbol{\bar \Delta_\delta}$.    
\end{proof}
Using the set defined in~\eqref{eq:M_defn}, we can leverage a scalar multiplier $\Lambda$, facilitating a reduction in the dimensionality of the optimization problem for controller design. The resulting set $\boldsymbol{\bar{\Delta}}_\delta$ has a standard structure, and hence in this case the multiplier and robust analysis follow established formulas~\cite{scherer2000robust}. 
\ifbool{arxiv}{A constructive optimization problem to obtain a matrix $D$ that reduces conservatism can be found in Appendix~\ref{subsec:approx_param_unc_set}.}{}

\textit{Discussion:} Recent years have witnessed an increasing interest in designing feedback controllers robust to parametric uncertainties arising from system identification. Studies such as~\cite{berberich2022combining, van2020noisy} explored the design of stabilizing state-feedback controllers for systems with bounded energy disturbances, directly using data. Building on these foundations, \cite{berberich2022combining} further integrated prior knowledge on disturbances and system matrices into the design process. A common limitation of these methods is their inability to handle systems with measurement noise and their assumption that process disturbances are bounded. Instead, we propose a principled indirect approach for synthesizing data-driven robust controllers for systems with unbounded measurement noise, explicitly tailored to the uncertainty set derived in Section~\ref{sec:unc_quant}.

In contrast,~\cite{umenberger2019robust} proposed a method to synthesize robust controllers for systems with state measurements and Gaussian noise. This strategy provides an over-approximation of the uncertainty with a structure as in~\eqref{eq:bar_Delta}. Hence, by applying Prop.~\ref{prop:alternative_unc_set}, we can obtain a similarly simple set for systems with (noisy) output measurements and structural constraints. Furthermore, in the special case where noise-free state measurement are available, we recover the uncertainty set from~\cite{umenberger2019robust} using the proposed method (cf.~\ifbool{arxiv}{Appendix~\ref{subsec:approx_param_unc_set}}{Prop.~\ref{prop:alternative_unc_set}}). In this regard, the proposed uncertainty set in Prop.~\ref{prop:alternative_unc_set} extends~\cite{umenberger2019robust} to accommodate systems with measurement noise and accommodates integration of known structural constraints while recovering the same controller in the case of perfectly measured states.

\section{Predictive Control}\label{sec:pred_control}
In this section, we derive a predictive controller that can additionally account for probabilistic constraints. Specifically, we consider the following stochastic optimal control problem with $\Pi = \{\pi_t\}_{t=0}^\infty$ denoting the sequence of control laws:
\begin{subequations}\label{eq:infinite_horiz_soc}
    \begin{align}
    \min_{\Pi}\ \max_{\vartheta\in\Theta}  &\lim_{T\rightarrow\infty} \frac{1}{ T}\E\left[\sum_{t=0}^{T-1} \| x_{t}\|_{Q_\mathrm{c}}^2+\| u_{t}\|_{R_\mathrm{c}}^2\right]\label{eq:soc_cost}\\
    \mathrm{s.t.}\  & x_{t+1} = A(\vartheta)x_t + B(\vartheta)u_t + E w_t,  \label{eq:soc_dyn_proc}\\
    &y_t = Cx_t + v_t,\ \forall\vartheta\in\Theta_\delta, \label{eq:soc_dyn_meas}\\
    &\Pr\left(h_{j}^\top \begin{bmatrix} x_t \\ u_t \end{bmatrix} \leq 1\right)\geq p_{j},\  \forall j\in\I{1,r}, \label{eq:chance_constr}\\
    &w_t \sim \mathcal{N}(0, Q(\eta)),\  v_t \sim \mathcal{N}(0, R(\eta)),\\
    &x_0 \sim \mathcal{N}(\mu_{\mathrm{x}, 0}, \Sigma_{\mathrm{x},0}), \label{eq:init_state_soc}\\
    &u_t = \pi_t(\{y_i\}_{i=0}^{t-1}, \{u_i\}_{i=0}^{t-1})\end{align}
\end{subequations}
 We consider chance constraints~\eqref{eq:chance_constr} with a probability level $p_{j} \in (0, 1)$. Due to unbounded Gaussian disturbances and measurement noise, deterministic constraint satisfaction is not possible and instead chance constraints ensure that constraints are satisfied with a specified probability. The initial state is Gaussian distributed with known mean and variance~\eqref{eq:init_state_soc}. The objective of this problem is to minimize the expected cost~\eqref{eq:soc_cost}. By choosing the matrices $C_{\epsilon}$ and $D_{\epsilon}$, defined in eq.~\eqref{eq:perf_out}, such that $[ C_{\epsilon}^\top,\  D_{\epsilon}^\top]^\top [ C_{\epsilon},\ D_{\epsilon}] = \mathrm{diag}(Q_{\mathrm{c}}, R_{\mathrm{c}})$ with $R_{\mathrm{c}}\succ 0$, this cost is proportional to the the squared $\mathcal{H}_2$-norm of the channel $d \rightarrow \epsilon$ as in Sec.~\ref{sec:lqg}, while adhering to chance constraints~\eqref{eq:chance_constr}. By considering all parameters in the set $\Theta_\delta$, we can also provide probabilistic guarantees for the unknown true system, given Assumption~\ref{asm:param_set}. 

To provide a computationally tractable approach, we examine the affine output-feedback control strategy
\begin{align}\label{eq:mpc_control_law}
    x_{t+1}^\mathrm{c} &= A_\mathrm{c} x_t^\mathrm{c} + L y_t, \quad
    u_t = K x_{t}^\mathrm{c} + \nu_t, 
\end{align}
where $\nu_t$ is the optimized input in the MPC problem and $A_\mathrm{c}$, $K$, and $L$ correspond to the robust controller designed in Section~\ref{sec:lqg}. The parametrization in eq.~\eqref{eq:mpc_control_law} is chosen to optimize inputs $\nu_t$ for enforcing chance constraints while preserving the stability properties of the robust controller designed in Section~\ref{sec:lqg}. Similar to \eqref{eq:closed_loop_lfr}, incorporating the established feedback policy yields the closed loop dynamics:
\begin{equation}
    \xi_{t+1} = \mathcal{A}(\vartheta) \xi_{t} + \mathcal{B}_\nu(\vartheta) \nu_t + \mathcal{B}_\mathrm{d}d_t,
\end{equation}
where:
\begin{align}\label{eq:new_calAB}
    \mathcal{A}(\vartheta) &= \mathcal{\hat A} + \mathcal{B}_\mathrm{p} \Delta \mathcal{C}_\mathrm{q}, \
    \mathcal{B}_\mathrm{\nu}(\vartheta) = \mathcal{\hat B}_\mathrm{\nu} + \mathcal{B}_p \Delta J_\Delta \begin{bmatrix} 0 \\ I \end{bmatrix},\ \mathcal{\hat B}_\mathrm{\nu} = \begin{bmatrix} \hat{B} \\ 0 \end{bmatrix},
\end{align}
with $\Delta = I \otimes \tilde \vartheta^\top$ (cf. Lemma~\ref{lem:ab_param}). Now, we decompose the evolution of states into stochastic and nominal terms, as standard in SMPC frameworks~\cite{hewing2020recursively,arcari2023stochastic}. The nominal state $\xi_t^{\mathrm{z}} \in \mathbb{R}^{2n_\mathrm{x}}$  evolves according to the nominal dynamics:
\begin{equation}\label{eq:nom_dynamics}
    \xi_{t+1}^{\mathrm{z}} = \mathcal{A}(\vartheta)\xi_t^{\mathrm{z}} + \mathcal{B}_{\mathrm{v}}(\vartheta) \nu_t.
\end{equation}
The error state $\xi_t^{\mathrm{e}} :=\xi_t -\xi_t^\mathrm{z}$ satisfies:
\begin{equation}\label{eq:error_dyn}
    \xi_{t+1}^{\mathrm{e}} = \mathcal{A}(\vartheta)\xi_t^{\mathrm{e}} + \mathcal{B}_\mathrm{d} d_t.
\end{equation}
The initial conditions are given by:
\begin{equation}~\label{eq:xi_init}
    \xi_0^{\mathrm{z}} \sim \mathcal{N}(\mu_{\mathrm{\xi}, 0}, 0), \quad \xi_0^{\mathrm{e}} \sim \mathcal{N}(0, \Sigma_{\mathrm{\xi}, 0}),
\end{equation}
where:
\begin{equation}
    \mu_{\mathrm{\xi}, 0} = \begin{bmatrix} \mu_{\mathrm{x}, 0} \\  0\end{bmatrix} , \quad \Sigma_{\mathrm{\xi}, 0} = \begin{bmatrix} \Sigma_{\mathrm{x}, 0} &0 \\ 0 & 0 \end{bmatrix}.
\end{equation}
%
%
The proposed predictive control framework is derived in the following subsections. Section~\ref{subsec:tube} introduces a tube-based strategy to bound nominal dynamics for all $\vartheta \in \Theta_\delta$. Section~\ref{subsec:error} formulates a conservative estimate of the stochastic error covariance. Section~\ref{subsec:constr_tight} integrates nominal tubes and error covariance over-approximations to enforce chance constraints. The resulting MPC formulation is presented in Section~\ref{subsec:mpc}, followed by an analysis of its closed-loop properties in Section~\ref{subsec:cl_analysis}.

\subsection{Nominal Tube}\label{subsec:tube}
In this subsection, we leverage homothetic tubes to capture the evolution of the nominal augmented state $\xi^\mathrm{z}_t$ $\forall \vartheta\in \Theta_\delta$. 
Specifically, we construct a sequence of ellipsoidal sets, $\{\Xi_{t}\}_{t=0}^N$, spanning the prediction horizon, ensuring that $\xi^\mathrm{z}_t \in \Xi_{t}$. Particularly, these tubes are parameterized as:
\begin{equation}\label{eq:tube_defn}
    \Xi_{t} = \left\{ \xi \mid \|\xi-\bar{\xi}_{t}\|_  {\mathcal{P}} \leq \alpha_{t} \right\},
\end{equation}
centered around nominal trajectory predictions $\bar{\xi}_{t}$ following dynamics:
\begin{equation}\label{eq:pred_dyn}
    \bar{\xi}_{t+1} = \mathcal{\hat A} \bar{\xi}_{t} + \mathcal{\hat B}_\nu \nu_{t},
\end{equation}
starting from $\bar{\xi}_0 = \mu_{\xi, 0}$ and with scalings $\alpha_{t} \in \mathbb{R}_{\geq0}$, $\alpha_0 = 0$. The shape matrix $\mathcal{P}$ is designed offline to ensure compliance with the following assumption.
\begin{assumption}\label{asm:tube_design}
The shape matrix $\mathcal{P}$ is a common Lyapunov function with a known contraction rate $\rho \in (0,1)$; i.e.,
\begin{equation}\label{eq:contraction_rate}
    \mathcal{A}(\vartheta)^\top \mathcal{P} \mathcal{A}(\vartheta) \preceq \rho^2 \mathcal{P}, \quad \forall \vartheta \in \Theta_\delta.
\end{equation}
\end{assumption}    
Since the controller from Sec.~\ref{sec:lqg} ensures robust stability $\forall \vartheta\in\Theta_\delta$, this assumption is naturally satisfied with the Lyapunov certificate $\mathcal{P}=\mathcal{X}^{-1}$. A method  to compute a tailored contraction rate $\rho$ and shape matrix $\mathcal{P}$ can also be found in~\ifbool{arxiv}{Appendix~\ref{subsec:tube_design}}{\cite[Appendix~D.1]{balim2024data}}.  
The scaling parameters $\alpha_t$ are determined online to ensure $\xi^\mathrm{z}_t \in \Xi_{t}$ using the dynamics in the following proposition.
\begin{proposition}\label{prop:tube_dyn}
Let Asm.~\ref{asm:tube_design} hold, and consider dynamics in eq.~\eqref{eq:nom_dynamics},~\eqref{eq:pred_dyn} with an input sequence $\nu_t$, $t\in \N$, and
\begin{equation}\label{eq:tube_dyn}
     \alpha_{t+1}\geq\rho\alpha_{ t}+\left\|\left(\begin{bmatrix}
        \bar x_{t} \\ \bar u_{t}
    \end{bmatrix}^\top\otimes I\right)\Sigma_{\mathrm{J}, \vartheta, \delta}^{1/2}\right\|,
\end{equation}
with:
\begin{equation}\label{eq:bar_xu}
    \begin{bmatrix}
        \bar{x}_{t} \\ 
        \bar{u}_{t}
    \end{bmatrix} = \begin{bmatrix}
        I & 0 \\ 
        0 & K
    \end{bmatrix} \bar{\xi}_t + \begin{bmatrix}
        0 \\ 
        \nu_t
    \end{bmatrix}, \
    \Sigma_{\mathrm{J}, \vartheta, \delta} = (I \otimes \mathcal{P}^{1/2} \mathcal{B}_{\mathrm{p}}) J \Sigma_{\vartheta,\delta} J^\top (I \otimes \mathcal{P}^{1/2} \mathcal{B}_{\mathrm{p}})^\top. 
\end{equation}
Then, it holds that $\xi^\mathrm{z}_t \in \Xi_{t}$, $\forall \vartheta \in \Theta_\delta$, $\forall t\in\N$.
\end{proposition}
\begin{proof}
 The right-hand side of~\eqref{eq:tube_dyn} satisfies $\forall \xi_t \in \Xi_{t}$:
    \begin{align}
         &\rho\alpha_{ t}+\left\|\left(\begin{bmatrix}
        \bar x_{t} \\ \bar u_{t} \end{bmatrix}^\top\otimes I\right)\Sigma_{\mathrm{J}, \vartheta, \delta}^{1/2}\right\| 
        =\rho\alpha_{ t}+\max_{\|\tilde\vartheta\|\leq 1}\left\|\left(\begin{bmatrix}
        \bar x_{t} \\ \bar u_{t} \end{bmatrix}^\top\otimes I\right)\Sigma_{\mathrm{J}, \vartheta, \delta}^{1/2}\tilde\vartheta\right\| 
        =\rho\alpha_{ t}+\max_{\vartheta\in\Theta_\delta}\left\|\left(\begin{bmatrix}
        \bar x_{t} \\ \bar u_{t} \end{bmatrix}^\top\otimes I\right) \left(I \otimes \mathcal{P}^{1/2}\mathcal{ B}_\mathrm{p}\right)J\tilde\vartheta\right\| \notag\\
        =&\rho\alpha_{ t}+\max_{\vartheta\in\Theta_\delta}\left\|\mathcal{P}^{1/2}\mathcal{ B}_\mathrm{p}\left(\begin{bmatrix}
        \bar x_{t} \\ \bar u_{t} \end{bmatrix}^\top\otimes I\right)J\tilde\vartheta\right\| 
        =\rho\alpha_{ t}+\max_{\vartheta\in\Theta_\delta}\left\|\mathcal{ B}_\mathrm{p}\Delta J_\Delta \begin{bmatrix}
        \bar x_{t} \\ \bar u_{t} \end{bmatrix}\right\|_\mathcal{P} 
        \geq \max_{\vartheta\in\Theta_\delta}\| \mathcal{A}(\vartheta)(\xi_{t} -\bar\xi_{t})\|_\mathcal{P} + \left\|\mathcal{ B}_\mathrm{p}\Delta J_\Delta \begin{bmatrix}
        \bar x_{t} \\ \bar u_{t} \end{bmatrix}\right\|_\mathcal{P}\notag\\
        \geq&\max_{\vartheta\in\Theta_\delta}\left\| \mathcal{A}(\vartheta)(\xi_{t} -\bar\xi_{t}) + \mathcal{ B}_\mathrm{p}\Delta J_\Delta \begin{bmatrix}
        \bar x_{t} \\ \bar u_{t} \end{bmatrix}\right\|_\mathcal{P} 
        \stackrel{\eqref{eq:new_calAB}}{=}\max_{\vartheta\in\Theta_\delta} \| \mathcal{A}(\vartheta)(\xi_{t} -\bar\xi_{t}) + (\mathcal{A}(\vartheta) - \mathcal{\hat A})\bar \xi_t +  (\mathcal{B}_\mathrm{\nu}(\vartheta) - \mathcal{\hat B}_\mathrm{\nu}) \nu_{t}\|_\mathcal{P} \notag\\
        =&\max_{\vartheta\in\Theta_\delta}\| \mathcal{A}(\vartheta) \xi_{t} + \mathcal{B}_\mathrm{\nu}(\vartheta) \nu_{t} - \bar \xi_{t+1}\|_\mathcal{P}. 
    \end{align}
First, the definition of the maximum singular value is employed. Subsequently, the contraction rate defined in Asm.~\ref{asm:tube_design} is utilized. Using the tube containment condition~\eqref{eq:tube_defn} we showed that $\xi_{t+1}\in \Xi_{t+1}$, given $\xi_t \in \Xi_{t}$, for an arbitrary input $\nu_t$, $\forall \vartheta \in \Theta_\delta$. The claim can be extended to $\forall t\in\N$ by induction, since at $t=0$, $\bar \xi_0 \in \Xi_0$ for $\alpha_0=0$.
\end{proof}
The dynamics~\eqref{eq:tube_dyn} can be incorporated into a predictive controller framework as an LMI constraint. Next, we provide a method that establishes a conservative over-approximation to the derived dynamics which allows for a computationally cheaper formulation.
\begin{corollary}\label{cor:approx_tube_dyn}
    The properties in Proposition~\ref{prop:tube_dyn} remain valid if the LMI constraint~\eqref{eq:tube_dyn} is replaced by the following second-order cone constraint:
    \begin{equation}\label{eq:approx_tube_dyn}
        \alpha_{t+1} \geq \rho\alpha_{t} +\left\|\bar\Sigma^{1/2}_{\mathrm{J},\vartheta,\delta}\begin{bmatrix}
        \bar x_{t} \\ \bar u_{t} \end{bmatrix}\right\|,
    \end{equation}
    with $\bar x_t$, $\bar u_t$ as in~\eqref{eq:bar_xu}, and:
\begin{equation}
    \bar \Sigma_{\mathrm{J},\vartheta,\delta}=\sum_{i=0}^{2n_\mathrm{x}} (I \otimes e_{2n_\mathrm{x},i})^\top \Sigma_{\mathrm{J},\vartheta,\delta} (I \otimes e_{2n_\mathrm{x},i}),
\end{equation}
\end{corollary}
\begin{proof}
It suffices to show that the scaling parameters $\alpha_t$, obtained from~\eqref{eq:approx_tube_dyn}, provide an upper bound to those derived from~\eqref{eq:tube_dyn}.
\begin{align}
    &\left\|\bar\Sigma^{1/2}_{\mathrm{J},\vartheta,\delta}\begin{bmatrix}
        \bar x_{t} \\ \bar u_{t} \end{bmatrix}\right\|^2 
    =\sum_{i=0}^{2n_\mathrm{x}}\begin{bmatrix}
        \bar x_{t} \\ \bar u_{t} \end{bmatrix}^\top (I \otimes e_{2n_\mathrm{x},i})^\top \Sigma_{\mathrm{J},\vartheta,\delta} (I \otimes e_{2n_\mathrm{x},i})\begin{bmatrix}
        \bar x_{t} \\ \bar u_{t} \end{bmatrix} =\sum_{i=0}^{2n_\mathrm{x}}e_{2n_\mathrm{x},i}^\top\left(\begin{bmatrix}
        \bar x_{t} \\ \bar u_{t} \end{bmatrix} \otimes I\right)^\top\Sigma_{\mathrm{J},\vartheta,\delta}\left(\begin{bmatrix}
        \bar x_{t} \\ \bar u_{t} \end{bmatrix} \otimes I\right)e_{2n_\mathrm{x},i}\\
    =&\tr { \left(\begin{bmatrix}
        \bar x_{t} \\ \bar u_{t} \end{bmatrix}^\top \otimes I\right)\Sigma_{\mathrm{J},\vartheta,\delta}\left(\begin{bmatrix}
        \bar x_{t} \\ \bar u_{t} \end{bmatrix} \otimes I\right)}
    \geq\lambda_{\mathrm{max}}\left(\left(\begin{bmatrix}
        \bar x_{t} \\ \bar u_{t}
    \end{bmatrix}^\top\otimes I\right)\Sigma_{\mathrm{J}, \vartheta, \delta}\left(\begin{bmatrix}
        \bar x_{t} \\ \bar u_{t} \end{bmatrix} \otimes I\right)\right).\notag 
\end{align}
The last step uses the fact that trace of a symmetric positive semi-definite matrix is an upper bound to its maximum eigenvalue, applying square root to the first and last expression concludes the proof.    
\end{proof}
Note that $\bar \Sigma_{\mathrm{J},\vartheta,\delta}$ is computed offline and thus the LMI condition in~\eqref{eq:tube_dyn} is reduced to a second-order cone constraint (SOC). 

\subsection{Stochastic Error Tube}\label{subsec:error}
A common approach to address chance constraints is by pre-computing the variance of the stochastic error term during offline design~\cite{arcari2023stochastic,muntwiler2023lqg}. Since the parameter vector $\vartheta$ is uncertain, the following proposition provides an upper bound to the covariance matrix, considering the set $\Theta_\delta$, to satisfy the chance constraints~\eqref{eq:chance_constr}.
\begin{proposition}\label{prop:err_covar_bound}
    Consider any sequence of covariance matrices $\bar\Sigma_{\mathrm{\xi}, t}$, $t\in\N$, satisfying the following inequality:
    \begin{equation}\label{eq:err_covar_bound_condition}
        \mathcal{A}(\vartheta) \bar \Sigma_{\mathrm{\xi}, t} \mathcal{A}(\vartheta)^\top + \mathcal{B}_d \mathcal{B}_d^\top \preceq \bar \Sigma_{\mathrm{\xi}, t+1}, \  \forall t\in\N,\  \forall \vartheta \in \Theta_\delta,
    \end{equation}
    with $\bar \Sigma_{\mathrm{\xi}, 0} =  \Sigma_{\mathrm{\xi}, 0}$ according to~\eqref{eq:xi_init}. Then, the stochastic error dynamics~\eqref{eq:error_dyn} satisfy $\xi_t^\mathrm{e}\sim\mathcal{N}(0, \Sigma_{\mathrm{\xi},t})$ with $\bar \Sigma_{\mathrm{\xi},t}\succeq  \Sigma_{\mathrm{\xi},t}$, for any $\vartheta \in \Theta_\delta$ and $t\in \N$.
\end{proposition}
A suitable sequence of matrices $\bar\Sigma_{\mathrm{\xi}, t}$, $t\in\N$ can be computed offline through a semi-definite program (SDP), see~\ifbool{arxiv}{Appendix~\ref{subsec:error_covar_design}}{~\cite[Appendix~D.2]{balim2024data}} for details.

\begin{remark}
Given that the the $\mathcal{A}(\vartheta)$ is stable $\forall \vartheta \in \Theta_\delta$, the covariance matrices converge to a stationary upper bound. This stationary variance can be computed similarly. 
\end{remark}

\subsection{Constraint Tightening}\label{subsec:constr_tight}
In this section, we combine the effects of the stochastic error tube (Sec.~\ref{subsec:error}) and the homothetic tube (Sec.~\ref{subsec:tube}) to ensure satisfaction of the chance constraints~\eqref{eq:chance_constr}.
\begin{proposition}\label{prop:constraint_tightening}
    Suppose that Asm.~\ref{asm:param_set}--\ref{asm:tube_design} hold and $\bar \Sigma_{\xi,t}$ satisfies conditions from Prop.~\ref{prop:err_covar_bound}. Consider the dynamics~\eqref{eq:state-space}, control law~\eqref{eq:mpc_control_law}, and tube dynamics in Prop.~\ref{prop:tube_dyn} or Cor.~\ref{cor:approx_tube_dyn}. Suppose further that:
\begin{align}\label{eq:chance_constr_tightened}
    h_{j}^\top\begin{bmatrix}
        \bar x_t \\ \bar u_t
    \end{bmatrix} \leq 1 - c_{j, t} - \alpha_{t}f_{j},\ \forall j \in \I{1,r}
\end{align}
for all $t\in\N$, with $\bar x$, $\bar u$ from eq.~\eqref{eq:bar_xu}, and:
\begin{align}
    c_{j,t} &= \Phi^{-1}(p_{j}) \left\|\bar \Sigma_{\mathrm{\xi}, t}^{1/2}\begin{bmatrix}I &  0 \\ 0 & K\end{bmatrix}^\top h_{j}\right\|, \label{eq:chance_constr_tightening} \\
     f_{j} &= \left\|\mathcal{P}^{-1/2}\begin{bmatrix}I &  0 \\ 0 & K\end{bmatrix}^\top h_{j}\right\| \label{eq:nom_tightening_term},
\end{align}
where $\Phi^{-1}$ is the quantile function of the standard normal distribution. Then, the chance constraints~\eqref{eq:chance_constr} are satisfied.
\end{proposition}
\begin{proof}
    Since $\bar \Sigma_{\xi,t}$ satisfies the conditions in Prop.~\ref{prop:err_covar_bound}:
    \begin{equation}\label{eq:ct_intera}
        \Pr\left(h_{j}^\top \begin{bmatrix}
        I & 0 \\ 0 & K
    \end{bmatrix} \xi_t^{\mathrm{e}}\leq c_{j,t}\right)\geq p_{j},
    \end{equation}
    where $\Phi^{-1}$ is the quantile function of the normal distribution. Furthermore,  $\xi_t^{\mathrm{z}}\in \Xi_t$ with \eqref{eq:tube_defn} and \eqref{eq:nom_tightening_term} implies:
    \begin{equation}\label{eq:ct_interb}
        h_j^\top \begin{bmatrix}I & 0 \\ 0 & K\end{bmatrix}{\xi}_{t}^\mathrm{z} \leq h_j^\top \begin{bmatrix}I & 0 \\ 0 & K\end{bmatrix}{\bar\xi}_{t} + \alpha_j f_{j,t}.
    \end{equation}
    Note $\xi^\mathrm{e}_t$ is completely independent of the optimized input $\nu_t$ and nominal state $\xi^\mathrm{z}_t$ and thus,
    \begin{align}
        &\Pr\left(h_{j}^\top \begin{bmatrix}
        x_{t}\\ u_{t}
    \end{bmatrix} \leq 1\right)
    = \Pr\left(h_{j}^\top \left(\begin{bmatrix}
        I & 0 \\ 0 & K
    \end{bmatrix}(\xi_t^{\mathrm{e}}+\xi_t^{\mathrm{z}})+\begin{bmatrix}
        0 \\ \nu_t
    \end{bmatrix}\right)\leq 1\right)
    \stackrel{\eqref{eq:ct_interb}}{\geq} \Pr\left(h_{j}^\top \left(\begin{bmatrix}
        I & 0 \\ 0 & K
    \end{bmatrix}(\xi_t^{\mathrm{e}}+\bar\xi_t)+\begin{bmatrix}
        0 \\ \nu_t
    \end{bmatrix}\right)\leq 1 - f_{j,t}\right). 
    \end{align}
    Finally, inequalities~\eqref{eq:ct_intera} and~\eqref{eq:chance_constr_tightened} imply satisfaction of the chance constraints~\eqref{eq:chance_constr}. 
\end{proof}
\subsection{Proposed MPC Formulation}\label{subsec:mpc}
This section introduces the proposed MPC scheme and summarizes the online and offline computations of the proposed D2PC framework. At each time step $t \in \N$, the following optimization problem is solved:
\begin{subequations}\label{eq:final_control_prob}
\begin{align}
\min _{\substack{\nu_{\cdot \mid t},\\ \bar \xi_{\cdot \mid t},\\ \alpha_{\cdot \mid t}} }\  & \sum_{i=0}^{T-1}(\|\bar \xi_{i\mid t}\|_{Q_{\mathrm{\xi}, \mathrm{c}}}^2+\|\nu_{i\mid t}\|_{R_\mathrm{c}}^2)+ \|\bar \xi_{T\mid t}\|_{S_{\mathrm{\xi}, \mathrm{c}}}\\
\mathrm { s.t. }\ 
&\bar \xi_{i+1 \mid t} = \mathcal{\hat A} \bar \xi_{i\mid t} + \mathcal{\hat B}_{\mathrm{\nu}} \nu_{i \mid t}, \label{eq:nom_pred_dyn}\\
&\text{Tube dynamics: }\label{eq:mpc_tube_dyn} \
 \alpha_{i+1\mid t} \geq \rho\alpha_{i\mid t} +\left\|\bar\Sigma^{1/2}_{\mathrm{J},\vartheta,\delta}\begin{bmatrix}
        \bar x_{i\mid t} \\ \bar u_{i\mid t} \end{bmatrix}\right\|,  \\
&\text{Tightened constraints: }\label{eq:mpc_constr} \
  h_{j}^\top \left(\begin{bmatrix}
    I & 0 \\ 0 & K
\end{bmatrix} \bar \xi_{i\mid t} + \begin{bmatrix}
    0 \\ \nu_{i\mid t}
\end{bmatrix} \right)\leq 1 - c_{j, t+i} - \alpha_{i\mid t}f_{j}, \\
& \quad\quad\quad\quad\quad\quad \quad  \ \ \forall j\in\I{1, r},\ \forall i\in\I{0, T-1},\notag\\
& \text {Terminal constraint: }(
\bar \xi_{T \mid t}, \ \alpha_{T \mid t}
) \in \Omega, \label{eq:mpc_term}\\
&\text{Initial state constraint: } \alpha_{0 \mid t} = \alpha_{1 \mid t-1}^\star, \  \bar \xi_{0 \mid t} = \bar \xi_{1 \mid t-1}^\star. 
\end{align}    
\end{subequations}
The proposed control problem provides a computationally tractable approach to address the outlined stochastic infinite-horizon control problem~\eqref{eq:infinite_horiz_soc} by approximating it with a finite-horizon problem of length $T$. The solution of \eqref{eq:final_control_prob} provide the optimal trajectories for the nominal predictions $\bar{\xi}_{\cdot \mid t}^{\star}$, the control input $\nu_{\cdot \mid t}^{\star}$, and the tube size $\alpha_{\cdot \mid t}^{\star}$. Consequently, the applied control input is defined as $u_t = Kx^{\mathrm{c}}_t + \nu_{0\mid t}^{\star}$, as detailed in eq.~\eqref{eq:mpc_control_law}. The initial conditions for the tube size $\alpha_{0\mid t}$, and the nominal prediction $\bar{\xi}_{0\mid t}$, are set to the corresponding values from the previous time-step, i.e. $\alpha_{1\mid t-1}^{\star}$ and $\xi_{1\mid t-1}^{\star}$, similar to~\cite{hewing2020recursively, arcari2023stochastic}. Note that the posed MPC problem is a SOC problem. It can be adapted to instead incorporate the tube dynamics from Proposition~\ref{prop:tube_dyn} by altering equation~\eqref{eq:mpc_tube_dyn}, resulting in an SDP.

The stage cost is calculated using the input term $\nu_{\cdot \mid t}$ and the nominal predictions $\bar{\xi}_{\cdot\mid t}$, where $Q_{\xi, \mathrm{c}} = \mathrm{diag}(Q_{\mathrm{c}}, K^{\top} R_{\mathrm{c}} K)$. The stage cost is applied to the nominal prediction term and optimized input term, aligning with robust tube MPC methods~\cite{schwenkel2022model}. Consequently, the optimization problem \eqref{eq:final_control_prob} results in $\nu_{0\mid t}^\star=0$ if the robust controller from Sec.~\ref{sec:lqg} adheres to the chance constraints. The terminal set $\Omega \in \mathbb{R}^{2n_{\mathrm{x}}+1}$, and the terminal cost weight $S_{\xi,\mathrm{c}}$ are specified in the following assumption:
\begin{assumption}[Terminal Conditions]\label{asm:terminal}
    The terminal set $\Omega$ contains the origin in its interior and $\forall (\xi,\alpha) \in \Omega$ we have:
\begin{enumerate}[label=\alph*), ref=\alph*)]
    \item \label{item:pos_inv} positive invariance\footnote{This condition is sufficient for ensuring positive invariance for the tube dynamics in Prop.~\ref{prop:tube_dyn} and in Cor.~\ref{cor:approx_tube_dyn}.}: 
    \begin{equation}
        \left(
            \mathcal{\hat A} \xi, \ \rho\alpha+\left\|\bar{\Sigma}^{1/2}_{\mathrm{J}, \vartheta, \delta} \begin{bmatrix}
        I & 0 \\ 0 & K
    \end{bmatrix}\xi\right\|
        \right)\in \Omega,
    \end{equation}
    
    \item \label{item:constr_sat} constraint satisfaction: 
    \begin{align}
    h_{j}^\top\begin{bmatrix}I &  0 \\ 0 & K\end{bmatrix}\xi &\leq 1 - c_{j, t} - \alpha f_{j},\  \forall t\in\N,\ \forall j\in \I{1, r},
    \end{align}
    \item \label{item:cost_dec} terminal cost decrease: 
    \begin{align}
        \|\mathcal{\hat A} \xi\|_{S_{\mathrm{\xi}, \mathrm{c}}}^2 - \|\xi\|_{S_{\mathrm{\xi}, \mathrm{c}}}^2 \leq - \left\|\xi\right\|_{Q_{\xi, \mathrm{c}}}^2.
    \end{align}
\end{enumerate}
\end{assumption}
This assumption can be naturally satisfied with an ellipsoidal set $\Omega$ and $S_{\mathrm{\xi}, \mathrm{c}}$ according to the Lyapunov equation, see~\ifbool{arxiv}{App.~\ref{subsec:terminal}}{\cite[App.~D.3]{balim2024data}} for details. Having introduced all necessary components, we now summarize the overall online and offline computations of our framework D2PC: 
\begin{algorithm}[h]
\caption{Online Computation}
\label{algo:online}
\begin{algorithmic}[1]
\Statex \% Execute at every time $t\in \N$
\State Measure the output $y_t$.
\State Set $\alpha_{0 \mid t} = \alpha_{1 \mid t-1}^\star$, $\bar{\xi}^\star_{0 \mid t}=\bar{\xi}^\star_{1 \mid t-1}$.
\State Solve the optimization problem~\eqref{eq:final_control_prob}.
\State Apply the control input $u_t = Kx^\mathrm{c}_t + \nu_{0\mid t}^\star$.
\State Update the controller state $x^\mathrm{c}_{t+1} = A_\mathrm{c}x^\mathrm{c}_t + Ly_t$.
\State Set $t=t+1$ and go back to 1.
\end{algorithmic}
\end{algorithm}
\begin{algorithm}[h]
\caption{Offline Computation}
\label{algo:offline}
\begin{algorithmic}[1]
\State Estimate $\vartheta$, $\eta$ from data with GEM (Sec.~\ref{sec:sysid}).
\State Quantify uncertainty and construct set $\Theta_\delta$ (Sec.~\ref{sec:unc_quant}).
\State Design robust controller: $A_\mathrm{c}$, $K$, $L$ (Sec.~\ref{sec:lqg}).
\Statex \% Predictive controller offline design (\ifbool{arxiv}{cf. App.~\ref{app:offline-design}}{{cf.~\cite{balim2024data} Appendix~D}}):
\State Design tube shape $\mathcal{P}$ and contraction rate $\rho$.
\State Establish bounds for stochastic error covariance $\bar \Sigma_{\xi, t}$.
\State Compute the tightening terms $c_{j,t}$, $f_j$ (Sec.~\ref{subsec:constr_tight}).
\State Construct terminal set $\Omega$, compute terminal weight $S_{\xi, \mathrm{c}}$.
\State Initialize $\alpha_{1 \mid -1}^\star = 0$, $\bar\xi_{1 \mid -1}^\star = \mu_{\xi,0}$.
\end{algorithmic}
\end{algorithm}

\subsection{Theoretical Analysis}\label{subsec:cl_analysis}
Next, we analyze the closed-loop theoretical properties. We demonstrate that the proposed controller not only adheres to the specified chance constraints but also recovers the same average cost incurred by the robust controller outlined in Sec.~\ref{sec:lqg}.
\begin{theorem}
    Let Assumptions ~\ref{asm:param_set},~\ref{asm:tube_design},~\ref{asm:terminal} hold and assume the optimization problem~\eqref{eq:final_control_prob} is feasible at $t=0$. Furthermore, suppose that the robust controller verifies the conditions in Thm.~\ref{thm:feas_control} for some  $\gamma$. Then~\eqref{eq:final_control_prob} is feasible for all $t\in \N$, the chance constraints~\eqref{eq:chance_constr} are satisfied for all $t\in \N$, and the average expected cost is no larger than $\gamma^2$ for the resulting the closed-loop system; i.e. 
    \begin{equation}
        \lim_{N\rightarrow\infty} \frac{1}{ N}\E\left[\sum_{t=0}^{N-1} \| x_{t}\|_{Q_\mathrm{c}}^2+\| u_{t}\|_{R_\mathrm{c}}^2\right] \leq \gamma^2.
    \end{equation}
\end{theorem}
\begin{proof}
    \textit{Recursive feasibility:} The recursive feasibility of the optimization problem can be proved using induction. Assume that~\eqref{eq:final_control_prob} is feasible at time $t-1$, then define the following candidate solution at time $t$:
    \begin{align}
    & {{\tilde \nu}}_{i \mid t}= 
    \begin{cases}\nu_{i+1 \mid t-1}^{\star} & \text { for } i=0, \ldots, T-2 \\
    0 & \text { for } i=T-1\end{cases} \\
    & {{\tilde{\bar \xi}}_{i \mid t}}= \begin{cases}\bar \xi_{i+1 \mid t-1}^{\star} & \text { for } i=0, \ldots, T-1 \\
     \mathcal{\hat A} \xi_{T \mid t-1}^{\star} & \text { for } i=T\end{cases} \\
    & {\tilde{\alpha}}_{i \mid t}= 
    \begin{cases} \alpha_{i+1 \mid t-1}^{\star} & \text { for } i=0, \ldots, T-1 \\\multicolumn{2}{c}{\rho\alpha_{T\mid t-1}^\star+ \left\|\bar{\Sigma}^{1/2}_{\mathrm{J}, \vartheta, \delta} \begin{bmatrix}
        I & 0 \\ 0 & K
    \end{bmatrix}\xi_{T\mid t-1}^\star\right\|}  \\&\text { for } i=T.\end{cases}
    \end{align}
    This shifted sequence directly satisfies the tightened constraints~\eqref{eq:chance_constr_tightened} for all $i \in [0, T-2]$. According to the terminal set's constraint satisfaction condition (Asm.~\ref{asm:terminal} condition~\ref{item:constr_sat}), these constraints also hold at $t=T-1$. Positive invariance of the terminal set (Asm.~\ref{asm:terminal} condition~\ref{item:pos_inv}) ensures ${(\tilde {\bar{\xi}}_{T\mid t},\ \tilde \alpha_{T\mid t})} \in \Omega$. Consequently, the solution adheres to the constraints in the control problem (see eqs.~\eqref{eq:mpc_tube_dyn},~\eqref{eq:mpc_constr},~\eqref{eq:mpc_term}). This confirms the feasibility of the candidate solution, validating recursive feasibility.\\
    \textit{Chance constraint satisfaction:} Since the control problem \eqref{eq:final_control_prob} is feasible for all $t\in\N$, the tightened constraints are satisfied for all $t\in\N$. By Prop.~\ref{prop:constraint_tightening}, satisfying the tightened constraints \eqref{eq:mpc_constr} ensures chance constraint satisfaction, given the independence of the stochastic error $\xi_t^\mathrm{e}$ from the nominal state $\xi_t^\mathrm{z}$ and controller input $\nu_t$. \\
    \textit{Asymptotic average cost bound:} To establish an asymptotic bound for the average cost, we first demonstrate that the applied input by the predictive controller $\nu_t$ vanishes asymptotically. Subsequently, we {show} input-to-state stability (ISS) of the nominal state $\xi^{\mathrm{z}}$, employing an approach analogous to that presented in~\cite{schwenkel2022model}. Finally, we ascertain that the cost associated with the nominal state diminishes asymptotically, rendering the cost exclusively dependent on the error dynamics $\xi^{\mathrm{e}}$.\\
    Denote the objective function for the problem~\eqref{eq:final_control_prob} as $J_\mathrm{T}(\bar\xi_{\cdot\mid t}, \nu_{\cdot \mid t})$, and use the suboptimality of the feasible candidate solution:
    \begin{align}
        &J_\mathrm{T}(\bar\xi_{\cdot\mid t}^{\star}, \nu_{\cdot \mid t}^{\star}) - J_\mathrm{T}(\bar\xi_{\cdot\mid t-1}^{\star}, \nu_{\cdot \mid t-1}^{\star}) 
        \leq J_\mathrm{T}({\tilde{\bar\xi}_{\cdot\mid t}, \tilde\nu_{\cdot \mid t}}) - J_\mathrm{T}(\bar\xi_{\cdot\mid t-1}^{\star}, \nu_{\cdot \mid t-1}^{\star}) \\
        =& \|\bar\xi_{T\mid t-1}^\star\|_{\bar Q_{\mathrm{c}}}^2+ \|\nu_{T\mid t-1}^\star\|_{R_{\mathrm{c}}}^2-\|\bar\xi_{0\mid t-1}^\star\|_{\bar Q_{\mathrm{c}}}^2- \|\nu_{0\mid t-1}^\star\|_{R_{\mathrm{c}}}^2  
        + \|\mathcal{A}\bar\xi_{T\mid t-1}^\star\|_{S_{\xi, \mathrm{c}}}^2 - \|\bar\xi_{T\mid t-1}^\star\|_{S_{\xi, \mathrm{c}}}^2, \notag\\
        \stackrel{\text{Asm.}~\ref{asm:terminal}\ref{item:cost_dec}}{\leq}& -\|\bar\xi_{0\mid t-1}^\star\|_{\bar Q_{\mathrm{c}}}^2- \|\nu_{0\mid t-1}^\star\|_{R_{\mathrm{c}}}^2. \notag
    \end{align}
    Using a telescopic sum till $t=N\in\N$ yields:
    \begin{align}
        \sum_{t=0}^{N}(\|\bar\xi_t\|_{\bar Q_{\mathrm{c}}}^2+ \|\nu_t\|_{R_\mathrm{c}}^2) &\leq J_\mathrm{T}(\bar\xi_{\cdot\mid 0}^{\star}, \nu_{\cdot \mid 0}^{\star}) - J_\mathrm{T}(\bar\xi_{\cdot\mid T}^{\star}, \nu_{\cdot \mid T}^{\star}) 
        \leq J_\mathrm{T}(\bar\xi_{\cdot\mid 0}^{\star}, \nu_{\cdot \mid 0}^{\star}),
    \end{align}
    using non-negativity of the cost $J_\mathrm{T}$.\\
    Next, we derive a bound on $J_\mathrm{T}$ using a case distinction. Suppose that the initial state is inside the terminal set $(\bar \xi_0 ,\ 0) \in \Omega$. The terminal set's positive invariance under $\nu = 0$ ensures that $\{\nu_t=0\}_{t=0}^{T-1}$ is a feasible candidate solution. By iteratively applying terminal cost decrease condition in Asm.~\ref{asm:terminal}, one can show that $J_\mathrm{T}(\bar\xi_{\cdot\mid 0}^{\star}, \nu_{\cdot \mid 0}^{\star})\leq \|\bar \xi_0\|_{S_{\xi, \mathrm{c}}}^2$. Given that the origin is in the interior of $\Omega$, there exists a class $\mathcal{K}$ function $\upalpha_\beta$, such that for any feasible $\bar{\xi}_0$, $J_{\mathrm{T}}(\bar{\xi}_{\cdot\mid 0}^{\star}, \nu_{\cdot\mid 0}^{\star}) \leq \upalpha_\beta(\|\bar \xi_0\|)$~\cite[Prop. B.25]{rawlings2017model}. Using $R_\mathrm{c} \succ 0$, we have:
    \begin{equation}\label{eq:nu_bound}
        c_0 \sum_{t=0}^{N}\|\nu_t\|^2 \leq \sum_{t=0}^{N}(\|\bar\xi_t\|_{\bar Q_{\mathrm{c}}}^2+ \|\nu_t\|_{R_\mathrm{c}}^2)\leq \upalpha_\beta(\|\bar \xi_0\|),
    \end{equation}
    for some $c_0> 0$.\\
    Next, we utilize the contraction condition from Asm.~\ref{asm:tube_design} to show that the nominal state dynamics $\xi_t^\mathrm{z}$ are ISS with respect to $\nu_t$. For any $\tau\in \N$, $c_1 > 0$, $\vartheta \in \Theta_\delta$, the following inequalities hold:
\begin{align}\label{eq:fy_ineq}
    \|\xi_{\tau+1}^{\mathrm{z}}\|^2_\mathcal{P} \leq&(1+c_1)\|A(\vartheta)\xi_\tau^{\mathrm{z}}\|_\mathcal{P}^2 + \left(1+\frac{1}{c_1}\right)\|\mathcal{B}_\nu(\vartheta) \nu_\tau\|_\mathcal{P}^2 
    \stackrel{\text{eq.}~\eqref{eq:contraction_rate}}{\leq} (1+c_1)\rho^2\|\xi_\tau^{\mathrm{z}}\|_\mathcal{P}^2 + \left(1+\frac{1}{c_1}\right)\|\mathcal{B}_\nu(\vartheta) \nu_\tau\|_\mathcal{P}^2,
\end{align}
where we applied the Young's inequality.\\
We select $c_1>0$ such that $\rho_{\mathrm{c}} = \sqrt{1+c_1}\rho < 1$. Given that $\Theta_\delta$ is compact, there exists a constant $c_2 > 0$ such that:
\begin{align}
    \left(1 + \frac{1}{c_1}\right)\|\mathcal{B}_\nu(\vartheta) \nu_\tau\|_\mathcal{P}^2 \leq c_2 \|\nu_\tau\|^2.
\end{align}
Using constants $\rho_{\mathrm{c}}$, $c_2$, we can write:
\begin{align}
    \|\xi_{\tau+1}^{\mathrm{z}}\|^2_\mathcal{P} \leq \rho_{\mathrm{c}}^2\|\xi_\tau^{\mathrm{z}}\|_\mathcal{P}^2 + c_2\|\nu_\tau\|^2.
\end{align}
Multiplying this inequality with $\rho_{\mathrm{c}}^{2(t-\tau-1)}$, applying a telescopic sum from $\tau=0$ to $\tau=t-1$ yields:
\begin{equation}
    \|\xi_t^\mathrm{z}\|_\mathcal{P}^2 \leq \rho_{\mathrm{c}}^{
{2}t} \|\xi_0^\mathrm{z}\|_\mathcal{P}^2 + c_2 \sum_{\tau=0}^{t-1} \rho_{\mathrm{c}}^{2(t-\tau-1)} \|\nu_\tau\|^2.
\end{equation}
Summing this inequality from $t=0$ to $t=N$ and using the geometric series $\sum_{t=0}^N \rho_{\mathrm{c}}^2 \leq 1/(1-\rho_{\mathrm{c}}^2)$, we obtain:
\begin{align}\label{eq:xiz_bound}
    \sum_{t=0}^N \|\xi_t^\mathrm{z}\|_\mathcal{P}^2 &\leq \frac{1}{1-\rho_{\mathrm{c}}^2} \|\xi_0^\mathrm{z}\|_\mathcal{P}^2 + \frac{c_2}{1-\rho_{\mathrm{c}}^2}\sum_{t=0}^N\|\nu_{{t}}\|^2
    \stackrel{\eqref{eq:nu_bound}}{\leq}\frac{1}{1-\rho_{\mathrm{c}}^2} \|\xi_0^\mathrm{z}\|_\mathcal{P}^2 + \frac{c_2 /c_0}{1-\rho_{\mathrm{c}}^2}\upalpha_\beta(\|\bar \xi_0\|). 
\end{align}
Since $\mathcal{P}$ is positive-definite we can find a constant $c_3 > 0$ such that:
\begin{align}
\label{eq:xiz_bound_with_c3}
    \sum_{t=0}^N \|\xi_t^\mathrm{z}\|^2&\leq \frac{c_3}{1-\rho_{\mathrm{c}}^2} \|\xi_0^\mathrm{z}\|_\mathcal{P}^2 + \frac{c_2c_3/c_0}{1-\rho_{\mathrm{c}}^2}\upalpha_\beta(\|\bar \xi_0\|).
\end{align}
Now, we connect the derived inequalities to the expected average cost. Observe that:
\begin{align}
    &\E \left[\|x_t\|_{Q_{\mathrm{c}}}^2+\|u_t\|_{R_{\mathrm{c}}}^2\right] =\E \left[\left\|\begin{bmatrix}
       \xi_t \\ \nu_t
    \end{bmatrix}\right\|_{{\mathcal{Q}_{\mathrm{c}}}}^2\right], \quad 
     {\mathcal{Q}_{\mathrm{c}}} = \begin{bmatrix}
         I & 0 & 0 \\ 0 & K & I
     \end{bmatrix}^\top \begin{bmatrix}
         Q_{\mathrm{c}} & 0 \\ 0 & R_{\mathrm{c}}
     \end{bmatrix}\begin{bmatrix}
         I & 0 & 0 \\ 0 & K & I
     \end{bmatrix}.
\end{align}
Decompose the augmented state $\xi_t$ into the nominal part and error part:
\begin{align}
    \E \left[\|x_t\|_{Q_{\mathrm{c}}}^2+\|u_t\|_{R_{\mathrm{c}}}^2\right]= \E[\|\xi^\mathrm{e}_t\|^2_{Q_{\xi,\mathrm{c}}}] +\left\|\begin{bmatrix}
       \xi_t^{\mathrm{z}} \\ \nu_t
    \end{bmatrix}\right\|_{{\mathcal{Q}_{\mathrm{c}}}}^2,
\end{align}
where we utilized that $\xi^{\mathrm{e}}_t$ is independent of $\xi^{\mathrm{z}}_t$ and $\nu_t$ and it has zero mean. Next, we utilize the bounds~\eqref{eq:nu_bound}, \eqref{eq:xiz_bound} to derive a bound on the average cost incurred by the nominal dynamics:
\begin{align}
    &\frac{1}{N}\sum_{t=0}^N\left\|\begin{bmatrix}
       \xi_t^{\mathrm{z}} \\ \nu_t
    \end{bmatrix}\right\|_{{\mathcal{Q}_{\mathrm{c}}}}^2 
    \leq \frac{\bar \lambda_c}{N}\sum_{t=0}^N \|\xi_t^{\mathrm{z}}\|^2 + \| \nu_t\|^2  
    \stackrel{\eqref{eq:nu_bound}, \eqref{eq:xiz_bound_with_c3}}{\leq} \frac{\bar \lambda_c}{N} \left(\frac{c_3}{1-\rho_{\mathrm{c}}^2} \|\xi_0^\mathrm{z}\|_\mathcal{P}^2 + \left(\frac{c_2c_3/c_0}{1-\rho_{\mathrm{c}}^2}+\frac{1}{c_0}\right)\upalpha_\beta(\|\bar \xi_0\|\right),
\end{align}
with $\bar \lambda_c = \lambda_{\mathrm{max}}({\mathcal{Q}_{\mathrm{c}}})$. Consequently, as $N\rightarrow \infty$ the average cost related to the nominal dynamics vanishes; yielding
\begin{align}
    &\lim_{N\rightarrow \infty}\sum_{t=0}^N\frac{1}{N}\E \left[\|x_t\|_{Q_{\mathrm{c}}}^2+\|u_t\|_{R_{\mathrm{c}}}^2\right] 
    = \lim_{N\rightarrow \infty}\frac{1}{N}\sum_{t=0}^N\E[\|\epsilon\|^2] \stackrel{\text{Thm.~\ref{thm:feas_control}}}{\leq} \gamma^2.
\end{align}
Recall that the dynamics of the error term~\eqref{eq:error_dyn} coincides with the dynamics investigated in the Thm.~\ref{thm:feas_control}. Thus, the asymptotic average cost bound is below $\gamma^2$.
\end{proof}

\textit{Discussion:} The provided predictive control scheme addresses the joint challenges of stochastic disturbances, uncertain parameters, and partial measurements by integrating aspects of stochastic and robust MPC approaches. 

In~\cite{lorenzen2019robust}, a robust MPC scheme utilizing polytopic homothetic tubes and parameter sets for systems with state measurements is introduced. Instead, our approach leverages ellipsoidal tubes to create a scalable MPC framework suitable for high-dimensional problems. In~\cite{parsi2022scalable}, an ellipsoidal homothetic tube-based predictive control framework for systems with linear fractional representation is presented. However, their optimization problem involves LMI constraints, leading to significant computational overhead. Our proposed approach, utilizing SOC tube dynamics in Cor.~\ref{cor:approx_tube_dyn}, significantly reduces computational demand with minimal additional conservatism, as shown in a subsequent numerical example{, and can handle unbounded stochastic noise}.

In~\cite{hewing2020recursively}, a predictive control strategy for systems with unbounded stochastic noise, called indirect feedback, is proposed, wherein the state evolution is decomposed into a nominal term and stochastic error terms. This methodology has been extended to include parametric uncertainty~\cite{arcari2023stochastic} and to accommodate systems with output measurements~\cite{muntwiler2023lqg}. We adopt a strategy akin to that of~\cite{arcari2023stochastic}: we bound the covariance of the stochastic error robustly and bound the nominal error through homothetic tubes. 
{ In~\cite{arcari2023stochastic}, the authors assume a polytopic parametric uncertainty set; however, constructing such a set from stochastic data would be nontrivial and conservative. In contrast, the strength of our approach lies the integration of the data-driven identification scheme by designing a control framework that is tailored to the resulting uncertainty set.} 
We deviate from the indirect-feedback approaches~\cite{hewing2020recursively,arcari2023stochastic,muntwiler2023lqg}, since we minimize a nominal cost independent of the online measurements. This approach establishes stronger performance guarantees for systems with parametric uncertainties compared to~\cite[Thm. 2]{arcari2023stochastic}, and by following a strategy akin to~\cite{schwenkel2022model}, we inherit the stability properties of the robust controller.

Recently, there has been an increasing interest in direct data-driven approaches~\cite{coulson2021distributionally,yin2021maximum,yin2023stochastic, pan2022stochastic,breschi2023data,teutsch2026stochastic}. In~\cite{yin2022data, yin2023stochastic, coulson2021distributionally,teutsch2026stochastic}, direct data-driven methods are developed that ensures (open-loop) chance constraint satisfaction for stochastic systems. However, application requires additionally measurements of process noise or absence thereof. A common limitation these approaches share is that their guarantees are challenging to extend to closed-loop operation, and existing results in this direction are limited~\cite{berberich2024overview}. In contrast, our proposed approach can ensure chance constraint satisfaction, recursive feasibility, and establish performance guarantees for closed-loop operation with unbounded process and measurement noise, based on the derived parameter set (cf.~Asm.~\ref{asm:param_set}).

\section{Case Study: Chain of Mass-Spring-Damper System}\label{sec:example}
In the following, we demonstrate the complete pipeline of the proposed D2PC framework using a chain of mass-spring-damper systems and compare it with alternative approaches. All computations are carried out in Python on a server instance with an 8-core allocation from an AMD EPYC 9654 96-core processor and 24 GB RAM. The optimization problems were solved using MOSEK~\cite{mosek} for LMIs and ECOS~\cite{domahidi2013ecos} for SOCPs and QPs through the CVXPY interface~\cite{diamond2016cvxpy}. The implementation for the numerical example is available online: \url{https://github.com/haldunbalim/D2PC}.

\begin{figure}
\centering

\begin{minipage}[c]{0.50\columnwidth}
    \centering
    \includegraphics[width=\linewidth]{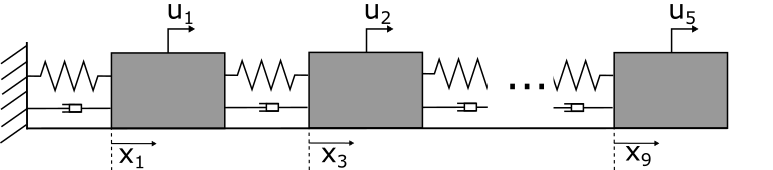}
\end{minipage}
\hfill
\begin{minipage}[c]{0.45\columnwidth}
    \centering
    \includegraphics[width=\linewidth]{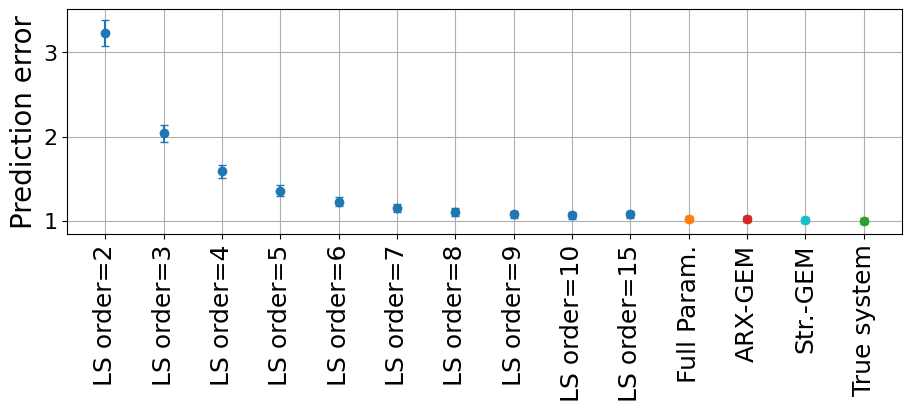}
\end{minipage}

\caption{
\textbf{Left:} Illustration for the spring-mass-damper system.
\textbf{Right:} Prediction error (normalized w.r.t.\ true system) across identified models and the true system. Error bars denote $\pm 3$ standard deviations. LS denotes models estimated using the least squares method, with the corresponding order.
}
\label{fig:smd-pred_err}
\end{figure}

\textit{Setup:} We consider a chain of $5$ mass-spring-dampers, see Fig.~\ref{fig:smd-pred_err}. The control input sets the forces on each mass separately, resulting in a system configuration where $n_\mathrm{x} = 10$ and $n_\mathrm{u}=5$. The system parameters are selected through uniform sampling: mass in the range $[0.9,\ 1.1] \, \text{kg}$, spring constant in the range $[1.8,\ 2.2] \, \text{N/m}$, and damping constant in the range $[0.9,\ 1.1] \, \text{kg/s}$. The system equations are discretized using the forward Euler method with a time step $0.1 \, \mathrm{s}$. The velocity of each mass is subject to a noise term, $n_\mathrm{w}=5$ with covariance $Q=3\cdot10^{-4}I$. We consider that only position measurements are available, $n_\mathrm{y}=5$, which are influenced by measurement noise with covariance $R=3\cdot10^{-4}I$.

\textit{Parameter Identification:} We estimate the covariance matrices $Q,\ R$ of the form $\lambda I$ resulting in $\eta\in\R^2$. The structure of matrices $A$ and $B$ is known, but the mass-spring-damping constants are unknown, resulting in $\vartheta \in \R^{23}$. Indicating that the coupling structure and how position changes based on velocity is known, while the parameters associated to the accelerations are all unknown and to be estimated. We generate a measurement-input sequence of length $T=2\cdot 10^3$ by applying randomly sampled inputs $u_t \sim \mathcal{N}(0, 4I)$.
We compare the following parametrization and identification methods:
\begin{enumerate}[label=(\arabic*.)]
    \item GEM with fully parameterized $A$, $B$ matrices ($\vartheta \in \R^{150}$).\label{item:id_a}
    \item GEM with ARX structure ($\vartheta \in \R^{100}$).\label{item:id_b}
    \item GEM with known structure ($\vartheta \in \R^{23}$).\label{item:id_c}
    \item Least-squares (LS) estimated ARX structure and varying order $o$ ($\vartheta \in \R^{o \cdot n_\mathrm{y}(n_\mathrm{y} + n_\mathrm{u})}$)\label{item:id_d}
\end{enumerate}
The computation times for the parameter estimation are:
~\ref{item:id_a}:~~$ 2020.5$\,s~\ref{item:id_b}:~$43.2$\,s~\ref{item:id_c}:~$6.14$\,s and~\ref{item:id_d}:~$2-10$\,ms for varying orders.
Evidently, imposing structural constraints reduces the offline computation time of GEM. To assess the prediction error performance, we sample $10^3$ validation trajectories from the true system, each of length $2\cdot 10^3$, and calculate the single-step prediction error conditioned on the previous time-steps for each model. The models estimated with GEM predict the next output using Kalman filter recursions. As seen in Fig.~\ref{fig:smd-pred_err}, the models estimated by the GEM algorithm are comparable with higher-order models estimated by least squares. Additionally, we note that estimating a system model with a high order would complicate the following controller design. For the remainder of this numerical example we will consider the method~\ref{item:id_c}.

\textit{Uncertainty Quantification:} Next, we assess the reliability of the uncertainty characterization outlined in Sec.~\ref{sec:unc_quant}. We generate $10^3$ input-output trajectories, each of length $2\cdot 10^3$, and use the GEM algorithm to estimate the system model. Following the procedure described in Sec.~\ref{sec:unc_quant}, we compute a confidence ellipsoid $\Theta_\delta$ and estimate $\Pr[\vartheta \in \Theta_\delta]$ empirically. Table~\ref{tab:unc_quant} presents the estimated probability that the true system parameters fall within these high-probability credibility regions. 
Our numerical results show that $\Pr[\vartheta\in\Theta_\delta]\approx \delta$ also with finite-samples.

\begin{table}[h]
\centering
\caption{Estimated probability values for true system parameters $\vartheta$ to be contained in the set $\Theta_\delta$ for varying probability levels $\delta$.\\}
\label{tab:unc_quant}
\begin{tabular}{c|ccccc}
\hline
 $\delta$ & 0.8 & 0.85 & 0.9 & 0.95 & 0.99 \\ \hline
 $\Pr[\vartheta\in\Theta_\delta]$&   0.801 & 0.855 &    0.900 &  0.949  & 0.995      \\
 \hline
\end{tabular}
\end{table}

\begin{figure*}
\begin{minipage}[c]{0.32\textwidth}
    \centering
    \includegraphics[width=\linewidth]{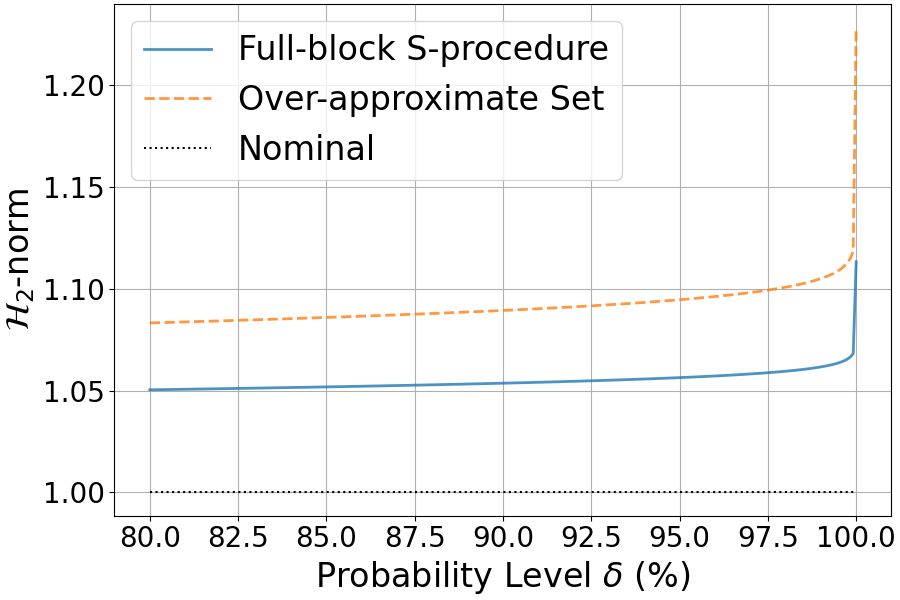}     
        \caption{Guaranteed closed-loop $\mathcal{H}_2$-norm vs. probability level $\delta$. Solid: proposed method with S-procedure (Lemma~\ref{lem:multiplier}), dashed: over-approximation (Prop.~\ref{prop:alternative_unc_set}), dotted: nominal LQG.}
    \label{fig:dynof_h2}                                 
    \end{minipage}
\hfill
    \begin{minipage}[c]{0.2915\textwidth}
        \centering
        \includegraphics[width=\linewidth]{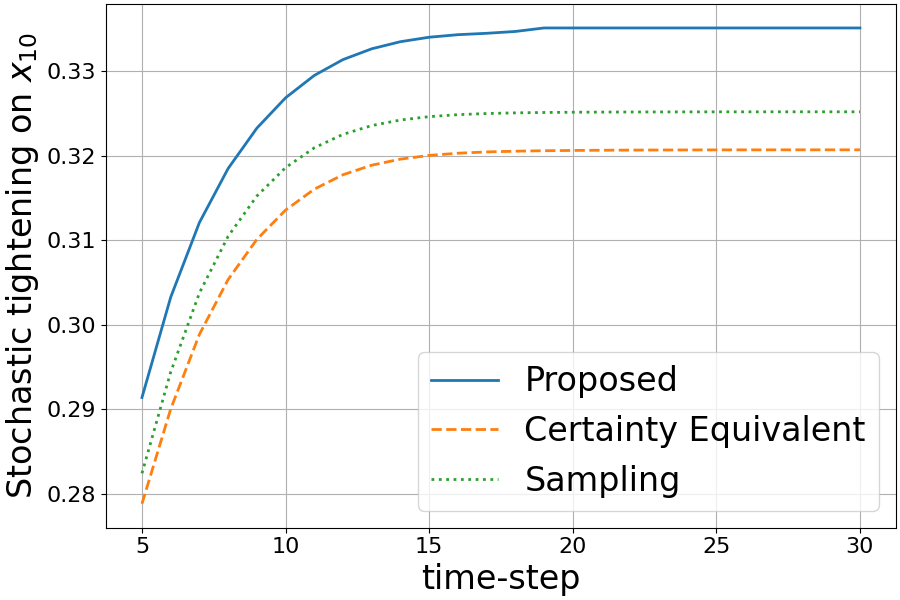}
        \caption{Constraint tightening $c_{j,t}$ using the proposed covariance bound (Prop.~\ref{prop:err_covar_bound}, blue), estimated $\hat \vartheta$ (orange), and sampled $\vartheta \in \Theta_\delta$ (green).}
        \label{fig:stoch_tight}
    \end{minipage}
\hfill
    \begin{minipage}[c]{0.32\textwidth}
        \centering
        \includegraphics[width=\linewidth]{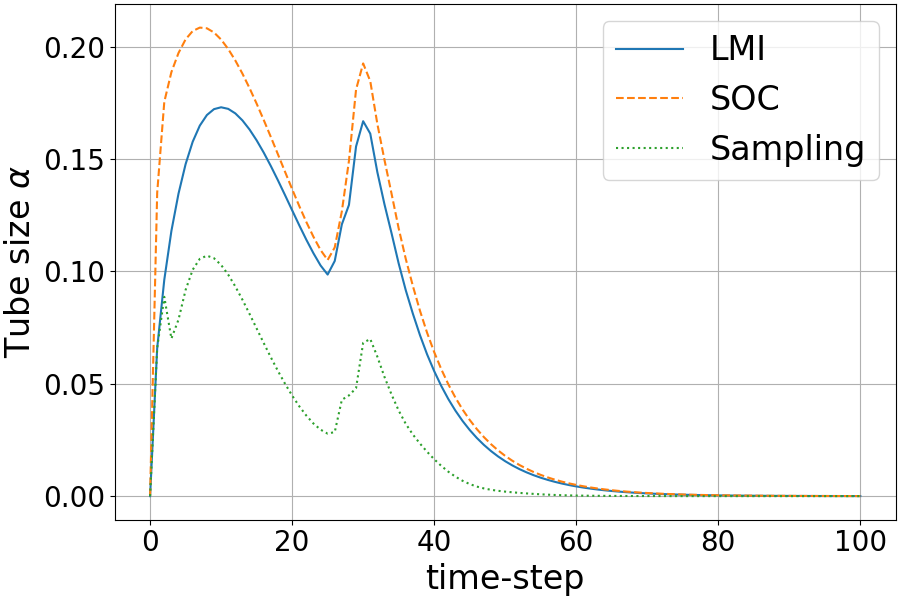}
        \caption{The evolution of nominal tube size $\alpha$ over time computed with LMI-based and SOC-based tube dynamics, and under-approximated using samples from $\vartheta \in \Theta_\delta$.}
        \label{fig:nom_tube}
    \end{minipage}
\end{figure*}

\begin{figure*}
\centering

\begin{minipage}[c]{0.58\textwidth}
    \centering
    \includegraphics[width=\linewidth]{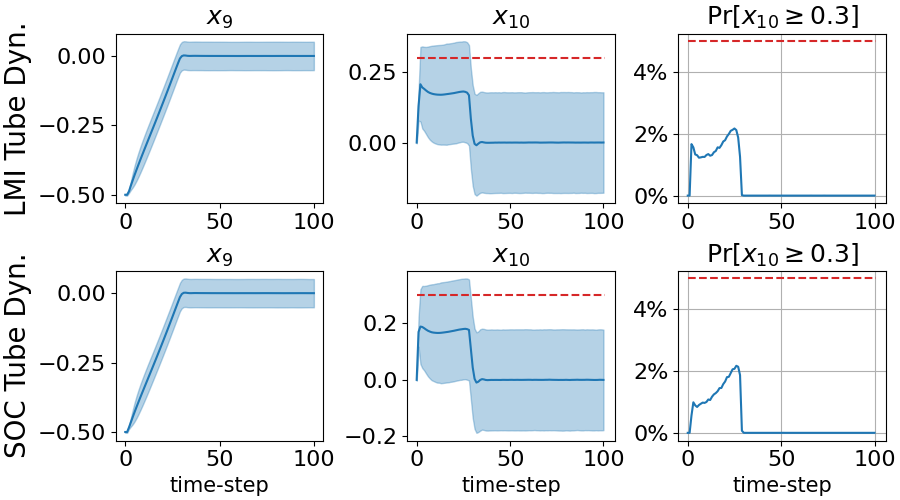}
\end{minipage}
\hfill
\begin{minipage}[c]{0.38\textwidth}
    \centering

    \begin{tabular}{cccc}\hline
    & \makecell{Cost}
    & \makecell{Comp. \\ Time (s)}
    & \makecell{Constr. \\ Violation (\%)} \\ \hline

    \makecell{Robust \\ Controller}
    & 1.000 & - & 100 \\ \hline

    \makecell{LMI \\ Tube Dyn.}
    & 2.189 & 1.396 & 2.2 \\ \hline

    \makecell{SOC \\ Tube Dyn.}
    & 2.309 & 0.029 & 2.2 \\ \hline

    \makecell{Nominal \\ SMPC}
    & 1.802 & 0.013 & 8.9 \\ \hline

    \end{tabular}

\end{minipage}

\caption{
\textbf{Left:} Position, velocity, and velocity constraint violation probability for the last mass, obtained from $10^5$ simulated trajectories. Shaded areas denote $\pm 3$ standard deviations. The first column corresponds to solving~\eqref{eq:final_control_prob} using the LMI tube dynamics (Prop.~\ref{prop:tube_dyn}), while the second uses the SOC tube dynamics (Cor.~\ref{cor:approx_tube_dyn}).
\textbf{Right:} Comparison of normalized cost, maximum empirical constraint violation probability across all time-steps, and average computation time for the considered controllers.
}
\label{fig:control_combined}

\end{figure*}

\textit{Robust Output-Feedback Controller Design:} Subsequently, we design output-feedback controllers, (Sec.~\ref{sec:lqg}), using cost matrices $C_\epsilon = [C,\ 0]$ and $D_\epsilon = [0,\ 10^{-4}I]$. The offline design takes, on average, $0.10$\,s for nominal LQG design, $5.77$\,s for robust controller design with full-block S-procedure, and $2.16$\,s for robust controller design with the approximate set. As seen in Fig.~\ref{fig:dynof_h2}, the simplified characterization (Prop.~\ref{prop:alternative_unc_set}) introduce small conservatism, while simplifying design to a scalar multiplier and thus reducing computational demand. 

\textit{Predictive Control:} In this section, we address a constrained control problem using the proposed framework. Specifically, we consider state and input chance constraints such that the velocity for each mass is bounded between $[-0.3, 0.3]$ and the inputs are bounded between $[-3.5, 3.5]$, each with probability $p_j=0.95$. The initial state distribution has a mean with each mass positioned at $-0.5$ with zero velocity and covariance $\Sigma_{\mathrm{x},0}=10^{-6}I$. The offline computation time to compute covariance bounds\ifbool{arxiv}{ using App.~\ref{subsec:error_covar_design}}{} with $N=19$ is $85.7$\,s, and to obtain the tube shape and contraction rate\ifbool{arxiv}{ using App.~\ref{subsec:tube_design}}{} was $99.7$\,s.

First, we investigate the stochastic tightening $c_{j,t}$ (Prop.~\ref{prop:constraint_tightening}) due to error covariances (Prop.~\ref{prop:err_covar_bound}), focusing specifically on the constraint concerning the last mass's velocity, see Fig~\ref{fig:stoch_tight}. To assess the conservatism of our approach, we replace the derived upper bound by the maximum covariance computed by using $10^4$ random samples from $\Theta_\delta \in \R^{23}$. By comparing the sampling-based estimate, we conclude that the proposed method over-approximates the true evolution with negligible conservatism.

Next, we compare the nominal tube size $\alpha$ using the tube dynamics proposed in Prop.~\ref{prop:tube_dyn} and Cor.~\ref{cor:approx_tube_dyn}, see Fig~\ref{fig:nom_tube}. For this purpose, we consider the inputs $\nu_t$ from an exemplary closed-loop trajectory. Similar to Fig.~\ref{fig:stoch_tight}, we under-approximate the maximal tube size using $10^4$ samples from $\vartheta \in \Theta_\delta$. The SOC-based tube dynamics results in a minimal increase in tube size. Compared to estimates based on sampling, both methods result in moderate conservatism. Conservatism may be due to the fact that the proposed tube propagation does not exploit time-invariance of the parameters.

Finally, we simulate $10^5$ trajectories using the proposed MPC~\eqref{eq:final_control_prob} with both tube dynamics with horizon $T=30$. Additionally, we implement a nominal SMPC scheme using parameters $\hat{\vartheta}$, neglecting the parametric uncertainty. The results are presented in Figure~\ref{fig:control_combined}. The nominal SMPC scheme fails to adhere to chance constraints, showing a violation probability over $5\%$. In contrast, the proposed framework with either tube dynamics consistently satisfies chance constraints across all time-steps. Although SOC-based tube dynamics provide a slightly worse performance, it reduces the computational complexity by a factor of over $40$.

\begin{figure*}[t]
\centering

\begin{minipage}[c]{0.45\textwidth}
    \centering
    \includegraphics[width=\linewidth]{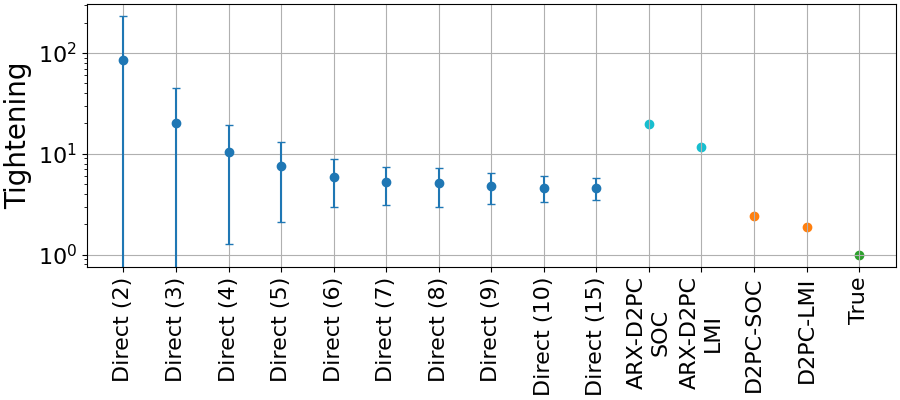}
\end{minipage}
\begin{minipage}[c]{0.45\textwidth}
    \centering

    \begin{tabular}{c|cc|cc|cc}
    \cline{2-7}
    & \multicolumn{2}{c|}{Direct (order)}
    & \multicolumn{2}{c|}{ARX-D2PC}
    & \multicolumn{2}{c}{D2PC} \\ \cline{2-7}
    & $2$ & $15$ & SOC & LMI & SOC & LMI \\ \hline
    Time (s) & $1.05$ & $1.00$
    & $0.09$ & $5.17$
    & $0.02$ & $1.32$ \\ \hline
    \end{tabular}

\end{minipage}

\caption{
\textbf{Left:} Comparison of tightening on the chance constraint on the last mass' position measurements for $t=30$ using the direct data-driven approach~\cite{yin2023stochastic} with varying orders and access to the true disturbances $w$, the proposed framework, and the true system parameters. Results are normalized by the tightening obtained using the true system parameters.
\textbf{Right:} Average computation times in seconds over $10$ independent trials for the direct data-driven method with different model orders and the proposed approach, using either SOC or LMI formulations with ARX or structured state-space models.
}
\label{fig:sddpc_combined}
\end{figure*}

\textit{Comparison with direct data-driven approach}:\footnote{We thank Mingzhou Yin for providing the implementation of the direct data-driven method.}

Lastly, we compare the conservatism and computational complexity of the proposed framework against the direct data-driven approach of~\cite{yin2023stochastic}; the results are shown in Fig.\ref{fig:sddpc_combined}. Since~\cite{yin2023stochastic} only accommodates chance constraints on the measurements $y_t$, we restrict the comparison to chance constraints imposed on the position measurements. The initial conditions are generated using the stationary Kalman filter associated with the true system, and we consider a step input sequence given by $u_t = 1_{n_\mathrm{u}}$. For the direct data-driven method, the tightening terms are computed using~\cite[Cor.11]{yin2023stochastic} by averaging over $10^4$ randomly generated initial output sequences obtained under zero input excitation. The tightening corresponding to the true system is obtained by propagating the initial state distribution through the error dynamics\eqref{eq:error_dyn}. For the proposed framework, the overall tightening is computed as the sum of the stochastic and nominal tube-dynamics contributions.

As illustrated in Fig.~\ref{fig:sddpc_combined}, the proposed approach yields significantly less conservative tightening when structural information is incorporated, while achieving comparable tightening values in the absence of structural assumptions. Moreover, the proposed SOC-based tube dynamics formulation exhibits substantially lower computational complexity than the direct data-driven method. It is important to emphasize that implementing the approach of~\cite{yin2023stochastic} requires access to the true disturbance realization sequence $w_t$ associated with the data. In practice, replacing the true disturbances with estimated values considerably degrades the reliability of the method. In contrast, the proposed framework only requires an upper bound on the disturbance covariance, making it substantially less restrictive and more practically applicable. Furthermore, to assess the computational complexity, we also solve both problems enforcing each position measurement to be in $[-1,\ 1]$ and input to be in $[-3.5,\ 3.5]$ for results. The optimization problem for the direct data-driven approach is solved by MOSEK, as ECOS fails to address this larger SOC problem.

This numerical example demonstrates that the proposed framework successfully addresses the control problem at hand. Furthermore, we demonstrate that our approach computes less conservative tightening terms compared to the direct approach, is less computationally demanding, and applicable with only input-output data.

\section{Conclusion}\label{sec:conclusion}

We present D2PC, a framework for designing reliable predictive controllers using stochastic input-output data. The framework encompasses four key elements: a method for parameter identification, a strategy for quantifying uncertainty in parameter estimates, an approach for designing robust dynamic output-feedback controllers tailored to the derived uncertainty set, and a predictive control scheme with closed-loop guarantees.  
The proposed framework bridges rigorous robust control and predictive control designs with the uncertainty estimates consistent with data-driven approaches. 
A broad open issue is improving the uncertainty quantification to provide rigorous finite-data bounds ensuring satisfaction of Assumption~\ref{asm:param_set} and relaxing the assumption of normal distributed noise. 

\subsubsection*{Conflicts of Interest}
The authors declare no conflicts of interest.
\subsubsection*{Data Availability Statement}
The code and data are available online: \url{https://github.com/haldunbalim/D2PC}
\bibliography{autosam}

\appendix
\section{Auxiliary Lemmas}\label{sec:aux_lemmas}
\begin{lemma}\label{lem:U_defn}
    Given an arbitrary matrix $V \in \mathbb{R}^{n \times m}$ with vectorization $v = \mathrm{vec}(V)$, it holds that:
    \begin{equation}\label{eq:unvec_rel}
        V = (I_n \otimes v^\top) (\mathrm{vec}(I_n) \otimes I_m).
    \end{equation}
\end{lemma}
\begin{proof}
    We demonstrate~\eqref{eq:unvec_rel} by comparing the $k$-th row of both sides for an arbitrary $k \in \I{1, \mathrm{n}}$. Consider the $k$-th row of the right-hand side of the equality:
    \begin{align}
        &e_{k,n}^\top (I_n \otimes v^\top) (\mathrm{vec}(I_n) \otimes I_m)
        = v^\top (e_{k,n}^\top \otimes I_{mn}) (\mathrm{vec}(I_n) \otimes I_m)
        = v^\top ((e_{k,n}^\top \otimes I_n) \otimes I_m) (\mathrm{vec}(I_n) \otimes I_m) \notag \\
        =& v^\top ((e_{k,n}^\top \otimes I_n) \mathrm{vec}(I_n) \otimes I_m)
        = v^\top (e_{k,n} \otimes I_m) 
        = e_{k,n}^\top V. 
    \end{align}
    This derivation confirms that the $k$-th row of both sides of the equality match for an arbitrary $k$.
\end{proof}

\begin{lemma}[{Adapted from \cite[Lemma 1]{berberich2022combining}}]\label{lem:Tdelta}
Let $M \in \R^{{n_\mathrm{m}}\times n_{\mathrm{w}}}$ be a full column-rank matrix. Then, $M\boldsymbol{\Delta}_{\mathrm{\delta}} = \boldsymbol{\tilde \Delta}_{\mathrm{\delta}}$ with $\boldsymbol{\Delta}_{\mathrm{\delta}}$ according to~\eqref{eq:param_set_delta} and
\begin{align}\label{eq:param_set_delta2}
    \boldsymbol{\tilde \Delta}_\delta =\Biggl\{\tilde \Delta \in \R^{n_\mathrm{m} \times n_\mathrm{w}n_\mathrm{\vartheta}} \:\Bigg|\: \begin{bmatrix}
         \tilde \Delta^\top \\ I_{n_\mathrm{m}} 
    \end{bmatrix}^\top \tilde P_{\Delta, \delta} \begin{bmatrix}
         \tilde \Delta^\top \\ I_{n_\mathrm{m}}
    \end{bmatrix}\succeq 0,  \notag \\ \forall \tilde P_{\Delta, \delta} \in \mathbf{\tilde P}_{\Delta, \delta} \Biggr\},\\
    \mathbf{\tilde P}_{\Delta, \delta} = \left\{\begin{bmatrix}
        -\Lambda \otimes \Sigma_{\vartheta,\delta}^{-1} & 0 \\ 0 & M\Lambda M^\top
    \end{bmatrix} \:\middle |\: 0 \preceq \Lambda \in \R^{n_\mathrm{w}\times n_\mathrm{w}} \right\}.\label{eq:PDelta2}
\end{align}
\end{lemma}

\ifbool{arxiv}{%
\section{Details for GEM Implementation}\label{sec:gem_deets}
This section details the E-step and GM-step of Algorithm~\ref{algo:em}.
\subsection{E-step}\label{sec:E}
The following proposition shows how to compute the conditional log-likelihood function $\mathcal{Q}(\theta, \theta^\prime)$.
\begin{proposition}[{Adapted from \cite[Lemma 3.1]{GIBSON20051667}}]
    For any $\theta \in \Theta$, the conditional log-likelihood function $\mathcal{Q}(\theta, \theta^\prime)$ satisfies:
    \begin{align}\label{eq:cond_ll}
    -2 \mathcal{Q} (\theta , \theta^{\prime}) &\propto  \tr{\Sigma_0^{-1} \E_{\theta^{\prime}}\left[\tilde x_0\tilde x_0^{\top} \mid Y_T\right]}/T  
    + \log \operatorname{det} \Sigma_{\mathrm{x},0}(\eta)/T + \log \operatorname{det} Q(\eta) +  \log \operatorname{det} R(\eta)  \\
     &\quad + \tr{Q(\eta)^{-1}\left[\Phi_{\mathrm{+}}-\Psi_{\mathrm{\varphi}} \Gamma^\top  -\Gamma \Psi_{\mathrm{\varphi}}^\top +\Gamma \Sigma_{\mathrm{\varphi}} \Gamma^\top\right]} \notag 
    + \tr{R(\eta)^{-1}\left[\Phi_{\mathrm{y}}-\Psi_{\mathrm{x}} C^\top-C\Psi_{\mathrm{x}}^\top+C \Sigma_{\mathrm{x}} C^\top\right]} 
    \end{align}
    where 
    \begin{align}\label{eq:avg-stats}
        \begin{bmatrix}
            \Phi_{\mathrm{+}} & \Psi_{\mathrm{+\varphi}} \\ \star & \Sigma_{\mathrm{\varphi}}
        \end{bmatrix} &= 
        \frac{1}{T}\sum_{t=0}^{T-1}\E\left[\begin{bmatrix}
            E^\dagger  x_{t+1}\\ \varphi_t
        \end{bmatrix}\begin{bmatrix}
            E^\dagger  x_{t+1}\\ \varphi_t
        \end{bmatrix}^\top \middle | Y_\mathrm{T}, \theta^\prime\right] &\
        \begin{bmatrix}
            \Phi_{\mathrm{y}} & \Psi_{\mathrm{xy}} \\ \star & \Sigma_{\mathrm{x}}
        \end{bmatrix} = 
        \frac{1}{T}\sum_{t=1}^{T}\E\left[\begin{bmatrix}
            y_t\\ x_t
        \end{bmatrix}\begin{bmatrix}
           y_t\\ x_t
        \end{bmatrix}^\top \middle | Y_\mathrm{T}, \theta^\prime\right]
    \end{align}
with $\tilde x_0 = x_0 - \bar x_0(\eta)$, $\varphi_t = [x_t^\top,\ u_t^\top]^\top$, $\Gamma = E^\dagger[A(\vartheta),\ B(\vartheta)]$.
\end{proposition}

\subsection{GM-step}\label{sec:GM}
We commence with a proposition that establishes a special case where the M step has an analytical global maximizer.
\begin{proposition}[{Adapted from~\cite[Lemma 3.3]{GIBSON20051667}}]\label{prop:m-step}
    Suppose that there are no structural constraints on the model; i.e. $J=I$ and both $Q(\eta)$ and $R(\eta)$ are fully parameterized. Furthermore, let $\Sigma_{\mathrm{\varphi}}$, $\Sigma_{\mathrm{x}}$ as in~\eqref{eq:avg-stats} be positive-definite and consider $\hat \vartheta$, $\hat \eta$ according to:
    \begin{align}\label{eq:cf-mstep}
        &[A(\hat \vartheta),\  B(\hat \vartheta)] =E \Psi_{+\mathrm{\varphi}} \Sigma_{\mathrm{\varphi}}^{-1} + [A_0,\ B_0],\quad
         Q(\hat \eta) = \Phi_{\mathrm{+}} - \Psi_{\mathrm{+\varphi}} \Sigma_{\mathrm{\varphi}}^{-1} \Psi_{\mathrm{+\varphi}}^\top,  \quad
        R(\hat \eta) = \Phi_{\mathrm{y}} - \Psi_{\mathrm{xy}} \Sigma_{\mathrm{x}}^{-1} \Psi_{\mathrm{xy}}^\top,  \\
        &\hat{\bar x}_0(\hat \eta) = \E[x_{0}\mid Y_\mathrm{T}, \theta^\prime], \notag \quad 
        \hat  \Sigma_{\mathrm{x},0}(\hat \eta) = \mathrm{Var}[x_{0} \mid Y_\mathrm{T}, \theta^\prime].\notag
    \end{align}
    Suppose that $\hat \theta = (\hat \vartheta, \hat \eta) \in \Theta$, the system~\eqref{eq:state-space} parameterized by $\vartheta^\prime$ is controllable and observable, and the input is persistently exciting, i.e., 
     $\sum_{t=1}^T u_tu_t^\top \succ 0$.
    Then, $\hat \theta$ is a unique global maximizer of $\mathcal{Q}(\theta, \theta^\prime)$.
\end{proposition}
In general, imposing a specific structure on the $[E^\dagger A, E^\dagger B]$, $Q(\eta)$ or $R(\eta)$ may preclude an analytical solution for the unique global maximizer to the conditional log-likelihood function. For an extensive analysis on the conditions under which the M-step admits a closed-form solution, see~\cite{nemesin2014robust}. In cases where a closed-form solution is unattainable, the maximization of $ \mathcal{Q}(\theta, \hat\theta_k) $ can be achieved through iterative optimization techniques. 

In the following, we detail the specific implementation of the GM step used in our code framework. For the parameter set $\Theta$, we consider $\vartheta$ and $\bar{x}_0(\eta)$ to reside within a compact hypercube, and require the covariance matrices $Q(\eta)$, $R(\eta)$, and $\Sigma_{\mathrm{x},0}$ to have eigenvalues between specified positive bounds. The set $\Theta$ can be chosen sufficiently large to ensure it is non-restrictive. For the covariance matrices $Q(\eta)$, $R(\eta)$, we consider the following structure:
\begin{align}\label{eq:QR_param}
    Q(\eta) &= \sum_{i=1}^{n_\mathrm{Q}} (\Pi_{i}^{\mathrm{Q}})^\top Q_i(\eta_{\mathrm{q},i}) \Pi_{i}^{\mathrm{Q}},\quad 
    R(\eta) = \sum_{i=1}^{n_\mathrm{R}} (\Pi_{i}^{\mathrm{R}})^\top R_i(\eta_{\mathrm{r},i}) \Pi_{i}^{\mathrm{R}}\notag,
\end{align}
where $ \{\Pi_{i}^{\mathrm{Q}}\}_{i=1}^{n_\mathrm{Q}} $, $ \{\Pi_{i}^{\mathrm{R}}\}_{i=1}^{n_\mathrm{R}} $ are orthogonal projectors corresponding to the blocks $ Q_i(\eta_{\mathrm{q},i}) $, $ R_i(\eta_{\mathrm{r},i}) $, and $\eta_{\mathrm{q},i}$, $\eta_{\mathrm{r},i}$ are distinct parts of the vector $\eta$. Regarding the block matrices $ Q_i(\eta_{\mathrm{q},i}) $ we consider three scenarios:
\begin{enumerate}\label{enum:q_cases}
    \item Known matrix configuration: $ Q_i(\eta_{\mathrm{q},i}) = Q_0 $, where $ Q_0 $ is a predefined symmetric positive-definite matrix.
    \item Proportional to a known matrix: $ Q_i(\eta_{\mathrm{q},i}) = \lambda Q_0 $, with optimized parameter $ \lambda \in \mathbb{R}_{>0} $ and a predefined symmetric positive-definite matrix $ Q_0 $.
    \item Completely unknown matrix structure: $Q_i(\eta_{\mathrm{q},i})$ is an optimized symmetric positive-definite matrix.
\end{enumerate}
Similarly, we consider the same structural constraints for each block $R_i(\eta_{\mathrm{r},i})$. The variables $\Sigma_{\mathrm{x},0}(\eta)$, $\bar x_0$ are considered to be fully parameterized by $\eta_{\mathrm{x}}$ which is independent from $\eta_{\mathrm{q},i}$, $\eta_{\mathrm{r},i}$.

The following algorithm details the proposed GM algorithm, which exploits the structure~\eqref{eq:QR_param}. Denote $\Gamma_i(\vartheta_i) = \Pi_i^{\mathrm{Q}}  [E^\dagger A(\vartheta),\ E^\dagger B(\vartheta)] $, where $\vartheta_i$ is the minimal sub-vector of $\vartheta$. Accordingly, we identify minimum number of projector groups $\{\{\Pi_j^{\mathrm{Q}}\}_{j=1}^{n_i}\}_{i=1}^{n_{\Pi}}$, ensuring that the $\Gamma_i(\vartheta_i)$ for different groups have disjoint sub-vectors $\vartheta_i$. This segmentation enables the decomposition of the conditional log-likelihood function $\mathcal{Q}(\theta, \theta^\prime)$ into distinct sub-objectives. 
If analytical solutions yield parameters outside the set $\Theta$, a projection onto $\Theta$ is required. Alternatively, a local minimum in $\Theta$ can be computed using L-BFGS.

\begin{algorithm}
\caption{$\mathtt{GM}$ Algorithm}
\label{algo:gm}
\begin{algorithmic}[1]
\State \textbf{Input:} Current parameters $\vartheta$, $\eta$, smoothed state distributions~\eqref{eq:avg-stats}.
 \State Compute $\eta_{\mathrm{x}}$ using Prop.~\ref{prop:m-step}.
 \State Compute $\{\eta_{\mathrm{r},i}\}^{n_\mathrm{R}}_{i=1}$ using~\cite[Sec.2.C]{nemesin2014robust}.
 \For{projector group $\{\Pi_j^{\mathrm{Q}}\}_{j=1}^{n_i}$}
 \Statex \% Determine $\{(\Gamma_j(\vartheta_j), Q_j(\eta_{\mathrm{q},j}))\}_{j=1}^{n_i}$
 \If{$n_i > 1$ and All $Q_j(\eta_{\mathrm{q},j})$ are fixed}
    \State Use least-squares to determine $\{\vartheta_j\}_{j=1}^{n_i}$.
\ElsIf{$n_i > 1$}
    \State Use L-BFGS to determine $\{(\vartheta_j, \eta_{\mathrm{q},j})\}_{j=1}^{n_i}$.
 \ElsIf{Analytical solution exists, see~\cite{nemesin2014robust}}
        \State Use analytical solution to determine $\{(\vartheta_j, \eta_{\mathrm{q},j})\}_{j=1}^{n_i}$.
\Else
        \State Use L-BFGS to determine $\{(\vartheta_j, \eta_{\mathrm{q},j})\}_{j=1}^{n_i}$.
 \EndIf
 \EndFor
 \State Recover $\vartheta$, $\eta$ from $\eta_{\mathrm{x}}$, $\{\eta_{\mathrm{r},i}\}^{n_\mathrm{R}}_{i=0}$, $\{\{(\vartheta_j, \eta_{\mathrm{q},j})\}_{j=1}^{n_i}\}_{i=1}^{n_{\Pi}}$.
 \State \Return $\vartheta$, $\eta$. 
\end{algorithmic}
\end{algorithm}

The proposed scheme conforms with the condition~\eqref{eq:cond_oracle_alg} and thus Prop.~\ref{prop:em_convergence} applies. Consequently, our algorithm ensures a monotone increase of the likelihood 
and convergence to a stationary point,


\textbf{Discussion:} 
The integration of a structural constraints into the EM algorithm was first explored by Kim and Taylor~\cite{kim1995restricted}, where the closed-form solutions in the M-step is replaced with a maximization by Newton's method. Similarly, Holmes et al.~\cite{holmes2013derivation} considered the integration of constraints for state-space model identification using the EM algorithm, where the M-step utilized a technique similar to block coordinate ascent. While any algorithm that guarantees a monotonic increase in the conditional log-likelihood is sufficient for convergence to a stationary point, empirical evidence suggests that the use of quasi-Newton methods can significantly accelerate this convergence~\cite{acceleration-em}. Motivated by these findings, we use the Limited-memory Broyden–Fletcher–Goldfarb–Shanno (L-BFGS) algorithm~\cite{fletcher2000practical} in our implementation. Furthermore, we utilize the analytical solutions to obtain a global maximizer with minimal computational load contingent on their applicability~\cite{nemesin2014robust}.

\section{Robust Dynamic-Output Feedback Controller Synthesis}\label{sec:lqg_synth}
Theorem~\ref{thm:feas_control} presents a matrix inequality for the synthesis of dynamic output-feedback controllers. However, the condition is nonlinear in the decision variables $\Lambda$, $\mathcal{X}$, and the controller. We adopt the standard procedure in the literature, \textit{D-K iteration}, to design the controller, which alternates between robust synthesis with fixed multipliers and robust analysis for a fixed controller~\cite{doyle1983synthesis}.

\textit{D-step:} The following SDP can be used to establish and upper bound to the $\mathcal{H}_2$-norm of the system~\eqref{eq:closed_loop_lfr} for a given controller:
\begin{subequations}\label{eq:robust_analysis}
    \begin{align}
        \min_{\Lambda, \mathcal{X}}\ & \tr{\mathcal{C}_\epsilon \mathcal{X} \mathcal{C}_\epsilon^\top}\\
        \mathrm{s.t.}&\quad \eqref{eq:stab_cond_nonlin}
    \end{align}
\end{subequations}
\textit{K-step:} Now, we derive a convex problem to synthesize a robust controller given the multiplier $\Lambda \in \mathbf{S}_{++}^{n_\mathrm{w}}$. We begin by parameterizing $\mathcal{X}$ and its inverse with $X, \hat{X}, Y, \hat{Y} \in \mathbf{S}_{++}^{n_x}$ and full-rank matrices $U, V \in \mathbb{R}^{n_x \times n_x}$, capitalizing on their symmetry:
\begin{equation}
    \mathcal{X} = \begin{bmatrix} X & U^\top \\ U & \hat{X} \end{bmatrix}, \quad  
    \mathcal{X}^{-1} = \begin{bmatrix} Y & V \\ V^\top & \hat{Y} \end{bmatrix}.
\end{equation}
Next, we introduce an auxiliary full-rank matrix $T \in \mathbb{R}^{2n_x \times 2n_x}$:
\begin{equation}
    T = \begin{bmatrix} I & Y \\ 0 & V^\top \end{bmatrix}.
\end{equation}
Utilizing $T$, we formulate the following matrices for synthesizing the controller:
\begin{align}\label{eq:params_synth}
    &\mathcal{X}T = \begin{bmatrix}
        X & I \\ U & 0
    \end{bmatrix}, \  T^\top \mathcal{X}T = \begin{bmatrix}
        X & I \\ I & Y
    \end{bmatrix},\ T^\top \mathcal{B}_\mathrm{p}\Lambda = \begin{bmatrix}
        E\Lambda \\ Y E\Lambda
    \end{bmatrix}, \
    \mathcal{C}_\epsilon\mathcal{X}T =\begin{bmatrix}
        (C_\epsilon\mathcal{X} + D_\epsilon M)^\top \\ C_\epsilon^\top
    \end{bmatrix}^\top,\ \mathcal{C}_{\mathrm{q}} \mathcal{X}T = J_\Delta\begin{bmatrix}
        X & I \\ M & 0
    \end{bmatrix},\notag\\
    &T^\top \mathcal{A}\mathcal{X}T = \begin{bmatrix}
        \hat A X + \hat BM & \hat A \\ S & Y\hat A + F C
    \end{bmatrix}, \ 
    T^\top\mathcal{B}_{\mathrm{d}}= \begin{bmatrix}
        EQ^{1/2} & 0 \\ YEQ^{1/2} & FR^{1/2}
    \end{bmatrix}, 
\end{align}
with auxiliary matrix variables
\begin{align}\label{eq:params_synth_reverse}
    U &= V^{-1} - V^{-1} Y X,\quad M = KU, \quad F = VL, \
    S = VA_{\mathrm{c}}U +  Y \hat A X + FCX + Y\hat BM.
\end{align}
Note that the controller can be recovered using Eq.~\eqref{eq:params_synth_reverse} by choosing an arbitrary full-rank matrix $V$. To establish a condition equivalent to Eq.~\eqref{eq:stab_cond_nonlin}, we multiply it by $\mathrm{diag}(T, I)$ from the left and its transpose from the right, applying Schur's complement thereafter, yielding:
\begin{equation}\label{eq:stab_cond}
    \begin{bmatrix}
        -T^\top \mathcal{X}T & 0 & T^\top \mathcal{A}\mathcal{X}T & T^\top \mathcal{B}_d &  T^\top \mathcal{B}_p\Lambda  \\
        \star &-\Lambda \otimes \Sigma_{\vartheta,\delta}^{-1}  & \mathcal{C}_q \mathcal{X}T & 0 & 0 \\
        \star & \star & -T^\top \mathcal{X}T & 0 & 0 \\
        \star & \star & \star & -I & 0 \\
        \star & \star & \star & \star & -\Lambda
    \end{bmatrix} \prec 0. 
\end{equation}
For the $\mathcal{H}_2$ norm objective, we posit a matrix $\mathcal{W}$ and the condition:
\begin{equation}\label{eq:perf_cond}
    \begin{bmatrix}
   \mathcal{W} & \mathcal{C}_\epsilon \mathcal{X}T \\
    \star & T^\top \mathcal{X}T
\end{bmatrix} \succeq 0.
\end{equation}
Using Schur's complement implies $\mathcal{W} \succeq\mathcal{C}_\epsilon \mathcal{X}\mathcal{C}_\epsilon^\top$ and thus $\tr{\mathcal{W}} \geq \tr{\mathcal{C}_\epsilon \mathcal{X}\mathcal{C}_\epsilon^\top}$. Resultantly, $\tr{\mathcal{W}}$ establishes a bound to $\gamma^2$ which is the bound on squared $\mathcal{H}_2$-norm for the channel $d\rightarrow \epsilon$, $\forall \vartheta \in \Theta_\delta$.
We thus propose the following convex problem:
\begin{subequations}\label{eq:synth_lqg}
\begin{align}
    \min_{X, Y, \mathcal{W}, M, F, S} & \  \tr{\mathcal{W}} \\
    \text{s.t.} & \  \text{\eqref{eq:stab_cond},~\eqref{eq:perf_cond}},
\end{align}
\end{subequations}
which yields the robust controller.
\textit{D-K Iteration:} As noted before we obtain the parameters for output-feedback controller by iterating between \textit{D-step} and \textit{K-step}. This alternation decreases the objective monotonously, and we terminate the process when the change in objective is desirably small. For the first iteration we initialize the controller with the nominal LQG solution using the system matrices derived from $\hat \vartheta$.
\subsection{Approximate Parametric Uncertainty Set}\label{subsec:approx_param_unc_set}
In this section, we discuss computation of the matrix $D$ for the over-approximation in Prop.~\ref{prop:alternative_unc_set}. In particular, we provide the following SDP:
\begin{subequations}\label{eq:D_opt}
\begin{align}
    \min_{D,M}\  & t \label{eq:D_opt_a} \\
    \text{s.t.} \quad & M \preceq tI, ,\\
   & M \succeq I, \label{eq:D_opt_b}, \\
   & M = \Sigma_{\vartheta, \delta}^{1/2} J^\top (D \otimes I) J \Sigma_{\vartheta, \delta}^{1/2}.
\end{align}
\end{subequations}
Minimizing $t$ minimizes $\lambda_{\mathrm{max}}(M)$; thereby reduces the size of the uncertainty set. The lower bound~\eqref{eq:D_opt_b} acts as a normalization.

Next, we discuss a special case in which this approximation reduces to the method in~\cite[Lemma 3.1]{umenberger2019robust}: Consider that system matrices are fully parameterized (i.e., $J=I$), and the covariance matrix of the parameters exhibits a specific structure, namely $\Sigma_{\vartheta} = D_{\vartheta} \otimes I$, for some matrix $D_{\vartheta}$. In this case, the minimizer to~\eqref{eq:D_opt} is given by $D = cD_{\vartheta}$, with $c \in \mathbb{R}_{>0}$, and the resulting uncertainty set is identical to that proposed in~\cite[Lemma 3.1]{umenberger2019robust}. 
\section{Offline Design for MPC}\label{app:offline-design}
\subsection{Tube Design}\label{subsec:tube_design}
In this subsection, we propose a convex optimization problem aimed at determining the shape of the nominal tube $\mathcal{P}$, given a specified rate of contraction.
\begin{proposition}\label{prop:tube_design}
Consider $\rho \in (0,1)$ and $\mathcal{X}_\mathrm{P} \in \mathbf{S}_{+}^{2n_\mathrm{x}}$ obtained by solving the following optimization problem:
\begin{subequations}\label{eq:tube_prob}
\begin{align}
    \min_{\mathcal{X}_\mathrm{P}, \Lambda, \gamma,\rho} &\sum_{i=1}^r \gamma_i \label{eq:tube_prob_a}\\
    \mathrm{s.t.} \ &  \begingroup 
        \setlength\arraycolsep{.9 pt}
        \begin{bmatrix}  \\  \\ \star \\  \\ \\
        \end{bmatrix}^\top
        \left[\begin{array}{cc|cc}
        - \rho^2\mathcal{X}_\mathrm{P} & 0 & 0 & 0\\
        0 & \mathcal{X}_\mathrm{P} & 0 & 0 \\
        \hline
        0 & 0 & -\Lambda \otimes \Sigma_{\vartheta,\delta}^{-1}  & 0 \\
        0 & 0 & 0 & \mathcal{B}_\mathrm{p}\Lambda\mathcal{B}_\mathrm{p}^\top
        \end{array}\right]
        \begin{bmatrix} I & 0  \\
            \mathcal{\hat A}^\top & \mathcal{C}_\mathrm{q}^\top \\
            \hline
            0 & I  \\
            I & 0 
        \end{bmatrix}\endgroup \prec 0, \label{eq:tube_prob_b}\\
    &\begin{bmatrix}
    \mathcal{X}_\mathrm{P} & \mathcal{X}_\mathrm{P} \begin{bmatrix}
        I & 0 \\ 0 & K
    \end{bmatrix} h_{i} \\ \star & \gamma_i
    \end{bmatrix} \succeq 0,\   \forall i\in\I{1,r},\label{eq:tube_prob_c}\\
    &\begin{bmatrix}
        (1-\rho)^2 I & \Sigma_{\mathrm{J},\delta}^{1/2} \left(\begin{bmatrix}
            I & 0 \\ 0 & K
        \end{bmatrix} \otimes \mathcal{B}_\mathrm{p}^\top\right) \\ \star & (I \otimes \mathcal{X}_\mathrm{P})
    \end{bmatrix} \succeq 0.\label{eq:tube_prob_d}
\end{align}
\end{subequations}
Then, $\mathcal{P} = \mathcal{X}_\mathrm{P}^{-1}$ and $\rho$ satisfy the conditions outlined in Asm.~\ref{asm:tube_design} and $f_i^2 \leq \gamma_i$, $\forall i\in\I{1,r}$ with $f_i$ as in~\eqref{eq:nom_tightening_term}.
\end{proposition}
\begin{proof}
    Similar to~\eqref{eq:stab_cond_nonlin}, the condition~\eqref{eq:tube_prob_b} implies that for all $\vartheta\in\Theta_\delta$, we have:
\begin{equation}
    \mathcal{A}(\vartheta)\mathcal{X}_\mathrm{P}\mathcal{A}(\vartheta)^\top \preceq \rho^2\mathcal{X}_\mathrm{P}.
\end{equation}
Using Dualization Lemma we obtain:
\begin{equation}
    \mathcal{A}(\vartheta)^\top\mathcal{P}\mathcal{A}(\vartheta) \preceq \rho^2\mathcal{P},
\end{equation}
which verifies Assumption~\ref{asm:tube_design}. Further applying Schur's complement to condition~\eqref{eq:tube_prob_c} yields:
\begin{equation}
    h_{i}^\top \begin{bmatrix} I & 0 \\ 0 & K \end{bmatrix}^\top \mathcal{X}_\mathrm{P} \begin{bmatrix} I & 0 \\ 0 & K \end{bmatrix} h_{i} \stackrel{\eqref{eq:nom_tightening_term}}{=} f_i^2\leq \gamma_i,
\end{equation}
which proves the latter claim.
\end{proof}
The optimization problem~\eqref{eq:tube_prob} is a SDP for a fixed contraction rate $\rho$. To determine the solution we conduct a line search over the contraction rate. Since $\gamma_i \geq f_i^2$ minimizing the objective~\eqref{eq:tube_prob_a} minimizes the squared sum of constraint tightening terms, due to the nominal tube size for a fixed $\alpha$.

Note that scaling the tube shape matrix $\mathcal{P}$ by any positive constant preserves the validity of Asm.~\ref{asm:tube_design}. To eliminate degenerate solutions we use the constraint~\eqref{eq:tube_prob_d}. In the following, we show that condition~\eqref{eq:tube_prob_d} establishes a bound on $\mathcal{P}$. Consider a scenario where $\|\bar \xi_t\| \leq 1$, $\alpha_0 \leq 1$, $\nu_t=0$. Furthermore, consider $\alpha_t\leq 1$ at an arbitrary time $t$. Applying Schur's complement to eq.~\eqref{eq:tube_prob_d}:
\begin{align}
    (1-\rho)^2 I&\succeq (\star) ^\top\left(\left(\begin{bmatrix}
            I & 0 \\ 0 & K
        \end{bmatrix}^\top \otimes \mathcal{P}^{1/2}\mathcal{B}_\mathrm{p}\right)\Sigma_{\mathrm{J},\delta}^{1/2, \top}\right),\\
        &\succeq (\star)^\top \left(\left(\bar\xi_t^\top\begin{bmatrix}
            I & 0 \\ 0 & K
        \end{bmatrix}^\top \otimes \mathcal{P}^{1/2}\mathcal{B}_\mathrm{p}\right)\Sigma_{\mathrm{J},\delta}^{1/2, \top}\right)
        = (\star)^\top \left(\left(\begin{bmatrix}
            \bar x_t \\ \bar u_t
        \end{bmatrix}^\top \otimes \mathcal{P}^{1/2}\mathcal{B}_\mathrm{p}\right)\Sigma_{\mathrm{J},\delta}^{1/2, \top}\right), \notag
\end{align}
where we used that $\|\bar \xi_t\|\leq 1 \implies \bar \xi_t\bar \xi_t^\top \preceq I$. Consequently:
\begin{align}
    1 &\geq \rho+\left\|\left(\begin{bmatrix}
            \bar x_t \\ \bar u_t
        \end{bmatrix}^\top \otimes \mathcal{P}^{1/2}\mathcal{B}_\mathrm{p}\right)\Sigma_{\mathrm{J},\delta}^{1/2, \top}\right\|
    \geq
    \rho \alpha_t + \left\|\left(\begin{bmatrix}
            \bar x_t \\ \bar u_t
        \end{bmatrix}^\top \otimes \mathcal{P}^{1/2}\mathcal{B}_\mathrm{p}\right)\Sigma_{\mathrm{J},\delta}^{1/2, \top}\right\|=\alpha_{t+1}.
\end{align}
Therefore, $\alpha_{t+1} \leq 1$ verifies the tube dynamics condition in Prop.~\ref{prop:tube_dyn}. Since the time $t$ was arbitrary and $\alpha_0\leq 1$, by induction $\alpha_t\leq 1$, $\forall t \in \N$. Resultantly, Problem~\eqref{eq:tube_prob} minimizes the constraint tightening for $\|\bar \xi\| \leq 1$. 
\subsection{Error Covariance}\label{subsec:error_covar_design}
In this section, we present a method to systematically compute a sequence of covariance matrices $ \{\bar\Sigma_{\mathrm{\xi}, t}\}_{t=0}^N $ that satisfies the condition specified in Prop.~\ref{prop:err_covar_bound}.
\begin{proposition}\label{prop:err_covar_bound_prob}
Consider the sequence of covariance matrices $ \{\bar\Sigma_{\mathrm{\xi}, t}\}_{t=0}^N $ for some $N\in\N$ obtained by solving the following optimization problem:
\begin{subequations}\label{eq:error_covar_prob}
\begin{align}
    &\min_{\substack{\bar \Sigma_{\mathrm{\xi},t},\\ \Lambda, \gamma}}\sum_{t=1}^N\sum_{i=1}^r \gamma_{i,t} \label{eq:error_covar_prob_a}\\
    &\begingroup 
        \setlength\arraycolsep{.2 pt}
        \begin{bmatrix}  \\  \\ \star \\  \\ \\
        \end{bmatrix}^\top
        \left[\begin{array}{cc|cc}
        \mathcal{B}_\mathrm{d}\mathcal{B}_\mathrm{d}^\top- \bar \Sigma_{\mathrm{\xi},t+1}& 0 & 0 & 0\\
        0 & \bar\Sigma_{\mathrm{\xi}, t} & 0 & 0 \\
        \hline
        0 & 0 & -\Lambda \otimes \Sigma_{\vartheta,\delta}^{-1}  & 0 \\
        0 & 0 & 0 & \mathcal{B}_\mathrm{p}\Lambda\mathcal{B}_\mathrm{p}^\top
        \end{array}\right] \begin{bmatrix} I & 0  \\
            \mathcal{\hat A}^\top & \mathcal{C}_\mathrm{q}^\top \\
            \hline
            0 & I  \\
            I & 0 
        \end{bmatrix}
        \endgroup 
        \prec 0,  \notag\\ 
        & \qquad\qquad\qquad\qquad\qquad\qquad\qquad \forall t\in\I{0,N-1} \label{eq:error_covar_prob_b}
        \\
    &\begingroup 
        \setlength\arraycolsep{.2 pt}
        \begin{bmatrix}  \\  \\ \star \\  \\ \\
        \end{bmatrix}^\top
        \left[\begin{array}{cc|cc}
        \mathcal{B}_\mathrm{d}\mathcal{B}_\mathrm{d}^\top- \bar \Sigma_{\mathrm{\xi},N}& 0 & 0 & 0\\
        0 & \bar\Sigma_{\mathrm{\xi}, N} & 0 & 0 \\
        \hline
        0 & 0 & -\Lambda \otimes \bar\Sigma_{\vartheta,\delta}^{-1}  & 0 \\
        0 & 0 & 0 & \mathcal{B}_\mathrm{p}\Lambda\mathcal{B}_\mathrm{p}^\top
        \end{array}\right] \begin{bmatrix} I & 0  \\
            \mathcal{\hat A}^\top & \mathcal{C}_\mathrm{q}^\top \\
            \hline
            0 & I  \\
            I & 0 
        \end{bmatrix}
        \endgroup 
        \prec 0,  \label{eq:error_covar_prob_c}\\
    &\begin{bmatrix}
    \bar\Sigma_{\mathrm{\xi}, t} & \Phi^{-1}(p_i)\bar\Sigma_{\mathrm{\xi}, t} \begin{bmatrix}
        I & 0 \\ 0 & K
    \end{bmatrix} h_{i} \\ \star & \gamma_{i,t}
    \end{bmatrix} \succeq 0,\   \forall i\in\I{1,r},\ \forall t\in\I{1,N}, \label{eq:error_covar_prob_d}\\
    &\bar\Sigma_{\mathrm{\xi}, 0} = \Sigma_{\mathrm{\xi}, 0}.
\end{align}    
\normalsize
\end{subequations}  
Then, $\bar \Sigma_{\xi,t} \succeq \Sigma_{\xi,t}$ $\forall t\in\I{0,N}$, $\bar \Sigma_{\xi,N} \succeq  \Sigma_{\xi,t}$  $\forall t\geq N$; i.e. $\{\bar\Sigma_{\mathrm{\xi}, t}\}_{t=0}^N$ verifies the condition Prop.~\ref{prop:err_covar_bound} and $\bar \Sigma_{\xi,N}$ can be used to bound the covariance of the stochastic error term for the time-steps $t \geq N$ and $c_{i,t}^2\leq \gamma_{i,t}$, $\forall i\in\I{1,r}$, $\forall t\in\I{1,N}$ with $c_{i,t}$ as in~\eqref{eq:chance_constr_tightening}.
\end{proposition}
\begin{proof}
Analogously to~\eqref{eq:stab_cond_nonlin}, the condition~\eqref{eq:error_covar_prob_b} dictates that for all $\vartheta\in\Theta_\delta$:
\begin{equation}
    \mathcal{A}(\vartheta) \bar \Sigma_{\mathrm{\xi}, t}\mathcal{A}(\vartheta)^\top + \mathcal{B}_\mathrm{d}\mathcal{B}_\mathrm{d}^\top\preceq \bar \Sigma_{\mathrm{\xi}, t+1},
\end{equation}
which implies~\eqref{eq:err_covar_bound_condition}. Thus, Prop.~\ref{prop:err_covar_bound} yields $\bar \Sigma_{\xi,t} \succeq \Sigma_{\xi,t}$. Similarly the condition~\eqref{eq:error_covar_prob_c} implies:
\begin{equation}
    \mathcal{A}(\vartheta) \bar \Sigma_{\mathrm{\xi},N}\mathcal{A}(\vartheta)^\top + \mathcal{B}_\mathrm{d}\mathcal{B}_\mathrm{d}^\top\preceq \bar \Sigma_{\mathrm{\xi}, N}.
\end{equation}
Based on or previous claim, we know that $\bar\Sigma_{\mathrm{\xi}, N} \succeq \Sigma_{\mathrm{\xi}, N}$. Furthermore, suppose that for some $t \geq N$, $\bar\Sigma_{\mathrm{\xi}, N} \succeq \Sigma_{\mathrm{\xi}, t}$, then:
\begin{align}
    \Sigma_{\mathrm{\xi}, t+1} &= \mathcal{A}(\vartheta)\Sigma_{\mathrm{\xi}, t}\mathcal{A}(\vartheta)^\top +\mathcal{B}_\mathrm{d}\mathcal{B}_\mathrm{d}^\top\\
    &\preceq \mathcal{A}(\vartheta)\bar\Sigma_{\mathrm{\xi}, N}\mathcal{A}(\vartheta)^\top +\mathcal{B}_\mathrm{d}\mathcal{B}_\mathrm{d}^\top
    \preceq \bar\Sigma_{\mathrm{\xi}, N}. \notag
\end{align}
Therefore, by induction $\bar \Sigma_{\xi,N} \succeq \Sigma_{\xi,t}$ $\forall t\geq N$. 
Applying Schur's complement to conditions~\eqref{eq:error_covar_prob_d} yields:
\begin{equation}
     (\Phi^{-1}(p_i))^2 h_{i}^\top \begin{bmatrix} I & 0 \\ 0 & K \end{bmatrix}^\top \Sigma_{\mathrm{\xi}, t} \begin{bmatrix} I & 0 \\ 0 & K \end{bmatrix} h_{i} = c_{i,t}^2 \leq \gamma_{i,t},
\end{equation}
thereby verifying the latter claim.
\end{proof}
The provided optimization problem is an SDP and, minimizing the objective~\eqref{eq:error_covar_prob_a} effectively reduces the squared sum of the constraint tightening due to the stochastic error term, analogously to the objective in~\eqref{eq:tube_prob_a}.
\subsection{Terminal Set Design}\label{subsec:terminal}
The following proposition introduces a terminal set similar to that described in~\cite{schwenkel2022model}, which satisfies Asm.~\ref{asm:terminal}.
\begin{proposition}
    Suppose Assumptions~\ref{asm:tube_design} hold and consider the following constants:
    \begin{align}\label{eq:underbar_c}
         \underbar c &= \min_{\substack{j\in\I{1,r},\\t\in\N}}(1 - c_{j, t}) / f_{j},
    \end{align}
    with $c_{j, t}$, $f_j$ as in~\eqref{eq:chance_constr_tightening},~\eqref{eq:nom_tightening_term} respectively and,
    \begin{equation}\label{eq:bar_sigma}
        \bar{\sigma} = \max_{\|\xi\|_\mathcal{P} \leq 1} \left\|\bar{\Sigma}^{1/2}_{\mathrm{J}, \vartheta, \delta} \begin{bmatrix}
        I & 0 \\ 0 & K
    \end{bmatrix}\xi\right\|.
    \end{equation}
    Furthermore, suppose that $\underbar{c}>0$. Then, the terminal set $\Omega = \{(\xi,\ \alpha) \mid \|\xi\|_\mathcal{P}+\alpha \leq \underbar c,\  \|\xi\|_\mathcal{P}\leq \frac{(1-\rho)}{\bar \sigma}\underbar c\}$ and $S_{\mathrm{\xi}, \mathrm{c}}$ as in~\eqref{eq:Sc_defn} satisfy Asm.~\ref{asm:terminal}.
\end{proposition}
\begin{proof}
    First, we show constraint satisfaction (Asm.~\ref{asm:terminal}~\ref{item:constr_sat}).
  For any $(\xi, \alpha) \in \Omega$, and any $t\in \N$, $j\in\I{1,r}$, it holds that 
  \begin{align}
    h_j^\top \begin{bmatrix}
            I & 0 \\ 0 & K
        \end{bmatrix}\xi+\alpha f_j \stackrel{\eqref{eq:nom_tightening_term}}{\leq } \|\xi\|_\mathcal{P}f_j+\alpha f_j
        {\leq} \underline{c} f_j \stackrel{\eqref{eq:underbar_c}}{\leq} 1-c_{j,t},
  \end{align}
  where the second inequality used the definition of the terminal set $\Omega$.
Next, we show the positive invariance of the terminal set:
    \begin{align}
        \underbar c &= \rho \underbar c + (1- \rho) \underbar c 
        \geq \rho\|\xi\|_\mathcal{P}+\rho\alpha +  \|\xi\|_\mathcal{P} \bar \sigma
        \geq \|\mathcal{\hat A}\xi\|_\mathcal{P}+\rho\alpha +  \left\|\bar{\Sigma}^{1/2}_{\mathrm{J}, \vartheta, \delta} \begin{bmatrix}
        I & 0 \\ 0 & K
    \end{bmatrix}\xi\right\|.  
    \end{align}
    Here, we used $\bar \sigma$ from eq.~\eqref{eq:bar_sigma} and the definition of contraction rate~\eqref{eq:contraction_rate}. Positive invariance,  Asm.~\ref{asm:terminal} condition~\ref{item:pos_inv}, can be ascertained using the last inequality and $ \frac{(1-\rho)}{\bar \sigma}\underbar c\geq \|\xi\|_\mathcal{P}\geq  \|\mathcal{\hat A}\xi\|_\mathcal{P} $.\\    
    Since $\hat \vartheta \in \Theta_\delta$, Asm.~\ref{asm:tube_design} implies that $\mathcal{\hat A}$ is Schur stable, there exists a unique $S_{\mathrm{\xi}, \mathrm{c}} \succ 0$ satisfying Lyapunov equation:
    \begin{equation}\label{eq:Sc_defn}
        \mathcal{\hat A}^\top S_{\mathrm{\xi}, \mathrm{c}}\mathcal{\hat A} -S_{\mathrm{\xi}, \mathrm{c}} = -\begin{bmatrix}
            Q_{\mathrm{c}} & 0 \\ 0 & K^\top R_{\mathrm{c}}K
        \end{bmatrix},
    \end{equation}
   i.e., the terminal cost condition (Asm.~\ref{asm:terminal}~\ref{item:cost_dec}) holds. $\hfill\square$
\end{proof}
Note that due to non-negativity of $\alpha$, the second condition in the definition of $\Omega$ is redundant if $1-\rho\geq \bar \sigma$. 
}{}
\end{document}